\pdfoutput=1
\documentclass[a4paper,fleqn,usenatbib]{mnras}
\usepackage{newtxtext,newtxmath}

\usepackage[T1]{fontenc}
\usepackage{ae,aecompl}
\usepackage{bm}

\usepackage[export]{adjustbox}
\usepackage{amsmath}	
\usepackage{amssymb}	
\usepackage{float}

\usepackage{multicol}
\usepackage{verbatim}   
\usepackage{threeparttable}
\usepackage{upgreek}

\newcommand{\lcdm}{$\Lambda$CDM}
\newcommand{\pcdm}{$\phi$CDM}
\newcommand{\om}{\Omega_{m0}}
\newcommand{\ol}{\Omega_{\Lambda}}
\newcommand{\ok}{\Omega_{k0}}
\newcommand{\FT}[1]{}

\title[Constraints from the QSO data]{Using quasar X-ray and UV flux measurements to constrain cosmological model parameters}

\author[N. Khadka, B. Ratra]{
 Narayan Khadka,$^{1}$\thanks{E-mail: nkhadka@phys.ksu.edu}
and Bharat Ratra,$^{1}$\thanks{E-mail: ratra@phys.ksu.edu}
\\
$^{1}$Department of Physics, Kansas State University, 116 Cardwell Hall, Manhattan, KS 66502, USA\\
}

\date{Accepted XXX. Received YYY; in original form ZZZ}

\pubyear{2019}

\hypersetup{draft}
\begin{document}
\label{firstpage}
\pagerange{\pageref{firstpage}--\pageref{lastpage}}
\maketitle

\begin{abstract}
Risaliti and Lusso have compiled X-ray and UV flux measurements of 1598 quasars (QSOs) in the redshift range $0.036 \leq z \leq 5.1003$, part of which, $z \sim 2.4 - 5.1$, is largely cosmologically unprobed. In this paper we use these QSO measurements, alone and in conjunction with baryon acoustic oscillation (BAO) and Hubble parameter [$H(z)$] measurements, to constrain cosmological parameters in six different cosmological models, each with two different Hubble constant priors. In most of these models, given the larger uncertainties, the QSO cosmological parameter constraints are mostly consistent with those from the BAO + $H(z)$ data. A somewhat significant exception is the nonrelativistic matter density parameter $\om$ where QSO data favor $\om \sim 0.5 - 0.6$ in most models. As a result, in joint analyses of QSO data with $H(z)$ + BAO data the one-dimensional $\Omega_{m0}$ distributions shift slightly toward larger values. A joint analysis of the QSO + BAO + $H(z)$ data is consistent with the current standard model, spatially-flat $\Lambda$CDM, but mildly favors closed spatial hypersurfaces and dynamical dark energy. Since the higher $\om$ values favored by QSO data appear to be associated with the $z \sim 2 - 5$ part of these data, and conflict somewhat with strong indications for $\om \sim 0.3$ from most $z < 2.5$ data as well as from the cosmic microwave background anisotropy data at $z \sim 1100$, in most models, the larger QSO data $\om$ is possibly more indicative of an issue with the $z \sim 2 - 5$ QSO data than of an inadequacy of the standard flat $\Lambda$CDM model.
\end{abstract}




\section{Introduction}
\label{sec:Introduction}
It is a well-established fact that the universe is now undergoing accelerated cosmological expansion. In general relativity, dark energy is responsible for the accelerated cosmological expansion. The simplest cosmological model consistent with this accelerated expansion is the spatially flat $\Lambda$CDM model, the current standard model (Peebles 1984).  In this model the accelerated expansion is powered by the time-independent and spatially homogenous cosmological constant ($\Lambda$) energy density. This model is consistent with many observations \citep{alam, Farooq2017, Scolnic2018, Plank2018} when dark energy contributes about $70\%$ of the current cosmological energy budget, approximately 25$\%$ contributed from cold dark matter (CDM), and the remaining 5$\%$ due to baryons. The standard model assumes flat spatial hypersurfaces.

While the $\Lambda$CDM model is consistent with many observations, it is based on the assumption of a spatially-homogeneous and time-independent dark energy density that is difficult to theoretically motivate. Additionally, data do not demand a time-independent dark energy density, and models in which the dark energy density decreases with time have been studied. In addition to the $\Lambda$CDM model, here we consider two dynamical dark energy models, the XCDM parametrization with a dynamical dark energy $X$-fluid and the $\phi$CDM model with a dynamical dark energy scalar field $\phi$.

While cosmological models with vanishing  spatial curvature are consistent with many observations, current observations do not rule out a little spatial curvature.\footnote{Discussion of observational constraints on spatial curvature may be traced through \cite{Farooq2015}, \cite{chen6}, \cite{Yu2016}, \cite{Rana2017}, \cite{Ooba2018a, Ooba2018b, Ooba2018c}, \cite{DESa}, \cite{Yu2018}, \cite{Park2018a, Park2018b, Park2018d, Park2018c, Park2019}, \cite{wei2018}, \cite{Xu2019}, \cite{Ruan2019}, \cite{Li2019}, \cite{Giambo2019}, \cite{Cole2019}, \cite{Eingorn2019}, \cite{Jesus2019}, \cite{Handley2019}, \cite{Wang2019}, \cite{Zhai2019}, \cite{Geng2020}, \cite{Kumar2020}, \cite{Efsta2020}, \cite{Di2020} and references therein.} So here, in addition to flat models, we also consider non-flat models with non-zero spatial curvature energy density. In this paper we test six different cosmological models, three spatially flat and three spatially non-flat.

These cosmological models have mostly been tested with data from low redshifts $z \sim 0$ up to redshift $z \sim 2.4$ baryon acoustic oscillation (BAO) measurements, as well as with cosmic microwave background (CMB) anisotropy data at $z \sim 1100$. They are poorly tested against data in the redshift range between $\sim 2.5$ and $\sim 1100$. To establish an accurate cosmological model and tighten cosmological parameter constraints, it is important to use additional cosmological probes, such as the quasar (QSO) flux - redshift data studied here. These QSO data probe the universe to $z \sim 5$ and are one of the few data sets that probe the $z \sim 2.5 - 5$ redshift range.\footnote{In the last decade or so, HII starburst galaxy data has reached to $z \sim 2.5$ \citep[and references therein]{Siegel2005, Mania2012, Gonzalez2019} while gamma ray burst data reach to $z \sim 8$ \citep[and references therein]{Lamb2000, SamushiaR2010, Demianski2019}.}

In 2015 Risaliti and Lusso published a systematic study that used quasar data to constrain cosmological parameters. The \cite{Risaliti2015} quasar sample has 808 quasar measurements extending over a redshift range $0.061 \leq z \leq 6.28$ which covers a significant part of the universe. These measurements have been used to constrain cosmological parameters \citep{Risaliti2015, Lopez2016, Lazkoz2019, Khadka2019} and the constraints obtained are consistent with those obtained from most other cosmological probes. However, the QSO data constraints \citep{Khadka2019} have larger error bars than those that result from BAO, Hubble parameter[$H(z)$], and some other data. This is because the empirical relation between the quasar's UV and X-ray luminosity, that is the basis of this method, has a large dispersion ($\delta = 0.32 \pm 0.008$). In 2019 Risaliti and Lusso enhanced these data by compiling a larger sample of quasars \citep{Risaliti2019}. For cosmological purposes, they selected 1598 quasars from a much larger number of sources. The dispersion of the $L_X - L_{UV}$ relation obtained from the new set of 1598 quasar measurements is smaller ($\delta = 0.23 \pm 0.004$) than that for the \cite{Risaliti2015} data. On the other hand, these new data give a relatively higher value of the matter density parameter in almost all models. This is one of the notable differences between the 2015 QSO and 2019 QSO data.

One major goal of our paper is to use the \cite{Risaliti2019} QSO data to constrain cosmological parameters in six cosmological models. We also consider how two different Hubble constant priors
affect cosmological parameter constraints. Since we use a number of different cosmological models here, we can draw somewhat model-independent conclusions about the QSO constraints. We find that the QSO measurements by themselves do not restrictively constrain cosmological parameters. However, given the larger error bars, the QSO constraints are mostly consistent with those that follow from the BAO + $H(z)$ observations, and when analyzed together the 2019 QSO measurements slightly tighten BAO + $H(z)$ data constraints in some of the models \citep[but less so than did the 2015 QSO data,][]{ Khadka2019}  and, more significantly, shift the matter density parameter ($\Omega_{m0}$) in most of the models to higher values. The QSO + BAO + $H(z)$ data are consistent with the standard spatially-flat $\Lambda$CDM model but mildly favor dynamical dark energy over a cosmological constant and closed spatial hypersurfaces over flat ones.

In most of the models we study here, the 2019 QSO data favor $\om \sim 0.5 - 0.6$. \cite{Risaliti2019} verify that the $z < 1.4$ part of the QSO data are consistent with $\om \sim 0.3$, which is also favored by most data up to $z \sim 2.5$, as well as by CMB anisotropy data at $z \sim 1100$, in most cosmological models. This 2019 QSO data preference for $\om \sim 0.5 - 0.6$ is therefore possibly more an indication of an issue with the $z \sim 2 - 5$ 2019 QSO data, and less an indication of the invalidity of the standard $\Lambda$CDM model \citep{Risaliti2019, Lusso2019}. Since the QSO data is one of the very few probes of the $z \sim 2 -5$ part of the universe, it is important to resolve this issue.

In Sec. 2 we summarize the models we use. In Sec. 3 we describe the data we use to constrain cosmological model parameters. In Sec. 4 we describe the techniques we use in our analyses. In Sec. 5 we compare 2019 QSO and 2015 QSO data constraints and present cosmological parameter constraints from the 2019 QSO data and the 2019 QSO + $H(z)$ + BAO data. We conclude in Sec. 6.
\section{Models}
\label{sec:models}
We use one time-independent and two dynamical dark energy models to constrain cosmological model parameters. We use flat and non-flat versions of each dark energy cosmological model and examine a total of six cosmological models. For dark energy we use a cosmological constant $\Lambda$ in the $\Lambda$CDM model, as well as an $X$-fluid dynamical dark energy density in the XCDM parametrization, and a scalar field $\phi$ dynamical dark energy density in the $\phi$CDM model.

In the $\Lambda$CDM model the redshift dependence of the Hubble parameter is
\begin{equation}
\label{eq:friedLCDM}
    H(z) = H_0\sqrt{\Omega_{m0}(1 + z)^3 + \Omega_{k0}(1 + z)^2 + \Omega_{\Lambda}},
\end{equation}
where $\om$ + $\ok$ + $\ol$ = 1. Here $\ol$ is the dark energy density parameter and $\om$ and $\ok$ are the current values of the non-relativistic matter and the spatial curvature energy density parameters. In the spatially-flat $\Lambda$CDM model we choose $\om$ and $H_0$ to be the free parameters while in the spatially non-flat $\Lambda$CDM model we choose $\om$, $\ol$, and $H_0$ to be the free parameters.

In the XCDM parametrization the dynamical dark energy density decreases with time. In this case dark energy is modeled as a fluid with equation of state $P_X$ = $\omega_{X}$ $\rho_{X}$. Here $P_X$ and $\rho_{X}$ are the pressure and energy density of the $X$-fluid, and $\omega_{X}$ is the equation of state parameter whose value is negative ($\omega_X < -1/3$). In this parametrization the Hubble parameter is
\begin{equation}
\label{eq:XCDM}
    H(z) = H_0\sqrt{\Omega_{m0}(1 + z)^3 + \Omega_{k0}(1 + z)^2 + \Omega_{X0}(1+z)^{3(1+\omega_X)}},
\end{equation}
where $\om$ + $\ok$ + $\Omega_{X0}$ = 1 and $\Omega_{X0}$ is the current value of the $X$-fluid energy density parameter. In the spatially-flat case we choose $\om$, $\omega_X$, and $H_0$ to be the free parameters while in the non-flat case we choose $\om$, $\ok$, $\omega_X$, and $H_0$ to be the free parameters. In the $\omega_{X}$ = $-1$ limit the XCDM parametrization becomes the $\Lambda$CDM model.

In the $\phi$CDM model a scalar field $\phi$ with potential energy density $V(\phi)$ provides the dynamical dark energy density that decreases with time \citep{peebles1988, Ratra1988, Pavlov2013}.\footnote{For discussions of observational constraints on the $\phi$CDM model see  \cite{chen2}, \cite{Samushia2007}, \cite{yashar2009}, \cite{SamushiaR2010}, \cite{Samushia2010}, \cite{chen4}, \cite{camp}, \cite{Farooq2013b}, \cite{Farooq2013a}, \cite{Avsa}, \cite{Sola2017}, \cite{Sola2018, Sola2019}, \cite{Zhai2017}, \cite{Sangwan2018}, \cite{Singh2019}, \cite{Mitra2019b}, \cite{Caok}, and references therein.} A commonly used $V(\phi)$ has the inverse power law form
\begin{equation}
\label{eq:phiCDMV}
    V(\phi) = \frac{1}{2}\kappa m_{p}^2 \phi^{-\alpha},
\end{equation}
with $\alpha$  a positive parameter, $m_{p}$ being the Planck mass, and 
\begin{equation}
\label{eq:kappa}
  \kappa = \frac{8}{3}\left(\frac{\alpha + 4}{\alpha + 2}\right)\left[\frac{2}{3}\alpha(\alpha + 2)\right]^{\alpha/2} .
\end{equation}
The equations of motion of this model are 
\begin{equation}
\label{field}
    \ddot{\phi} + \frac{3\dot{a}}{a}\dot\phi - \frac{1}{2}\alpha \kappa m_{p}^2 \phi^{-\alpha - 1} = 0,
\end{equation}
and
\begin{equation}
\label{friedpCDM}
    \left(\frac{\dot{a}}{a}\right)^2 = \frac{8 \uppi G}{3}\left(\rho_m + \rho_{\phi}\right) - \frac{k}{a^2}.
\end{equation}
Here $a$ is the scale factor, an overdot denotes a time derivative, $k$ is negative, zero, and positive for open, flat, and closed spatial geometries, the non-relativistic matter density is $\rho_m$ , and the scalar field energy density is
\begin{equation}
    \rho_{\phi} = \frac{m^2_p}{32\pi}[\dot{\phi}^2 + \kappa m^2_p \phi^{-\alpha}].
\end{equation}
The Hubble parameter in the $\phi$CDM model is
\begin{equation}
    H(z) = H_0\sqrt{\Omega_{m0}\left(1 + z\right)^3 + \Omega_{k0}\left(1 + z\right)^2 + \Omega_{\phi}\left(z, \alpha\right)},
\end{equation}
where
\begin{equation}
    \Omega_{\phi}(z, \alpha) = \frac{8 \uppi G \rho_{\phi}}{3H^2_0},
\end{equation}
with $G$ being the gravitational constant and $\om$ + $\ok$ + $\Omega_{\phi}(0, \alpha)$ = 1. In the $\phi$CDM model $\Omega_{\phi}(z,\alpha)$  has to be computed numerically. In the non-flat $\phi$CDM  model we choose $\om$, $\ok$, $\alpha$, and $H_0$ to be the free parameters while in  the spatially-flat $\phi$CDM model we choose $\om$, $\alpha$, and $H_0$ to be the free parameters. In the limit $\alpha\rightarrow0$ the $\phi$CDM model becomes the $\Lambda$CDM model.

\section{Data}
\label{sec:data}
The \cite{Risaliti2015} QSO compilation has 808 quasar flux-redshift measurements over a redshift range $0.061 \leq z \leq 6.28$. In this compilation most of the quasars are at high redshift, $\sim 77\%$ are at $z > 1$ and only $\sim 23\%$ are at $z < 1$. These data have a larger intrinsic dispersion ($\delta = 0.32 \pm 0.008$) in the $L_X - L_{UV}$ X-ray and UV luminosity relation which affects the error bars and so these data do not tightly constrain cosmological parameters. See \cite{Khadka2019} for cosmological constraints obtained from the 2015 QSO observations. 

To improve upon their 2015 data set, in 2019 Risaliti and Lusso published a compilation of 1598 quasars, chosen for the purpose of constraining cosmological parameters from a large sample of 7,237 sources \citep{Risaliti2019}.\footnote{We thank Elisabeta Lusso (private communication, 2019) for very kindly providing these data to us.} A significant portion of the QSOs in this new compilation are at lower redshift ($\sim$ 43$\%$ are at redshift $z \leq 1$), with QSOs in this new compilation distributed more uniformly over a smaller redshift range of $0.036 \leq z \leq 5.1003$ in comparison to the old data. The redshift distribution of the new quasar data is shown in Fig.\ 1. These QSOs have an $L_X - L_{UV}$ relation with a smaller intrinsic dispersion ($\delta = 0.23 \pm 0.004$). The main purpose of our paper is to use the 1598 QSO X-ray and UV flux measurements of \cite{Risaliti2019} to determine parameter constraints.\footnote{For cosmological parameter constraints derived from the 2019 QSO data, see \cite{Risaliti2019}, \cite{Lusso2019}, \cite{Melia2019}, \cite{yang2019}, \cite{Velten2020}, \cite{wei2020}, \cite{Lin2019}, \cite{Zheng2020}, and \cite{Mehrabi2020}.} We also compare the constraints from the 2019 QSO data to those that follow from the earlier \cite{Risaliti2015} QSO compilation.

Additionally, we compare the 2019 QSO data cosmological constraints to those computed from more widely used $H(z)$ measurements and BAO distance observations. The $H(z)$ and BAO measurements we use consist of 31 $H(z)$ observations over redshift $0.07 \leq z \leq 1.965$ and 11 BAO observations over redshift $0.106 \leq z \leq 2.36$. The $H(z)$ and BAO measurements we use are given in Table 2 of \cite{Ryan2018} and Table 1 of \cite{Ryan2019}.

\begin{figure}
    \includegraphics[width=\linewidth, right]{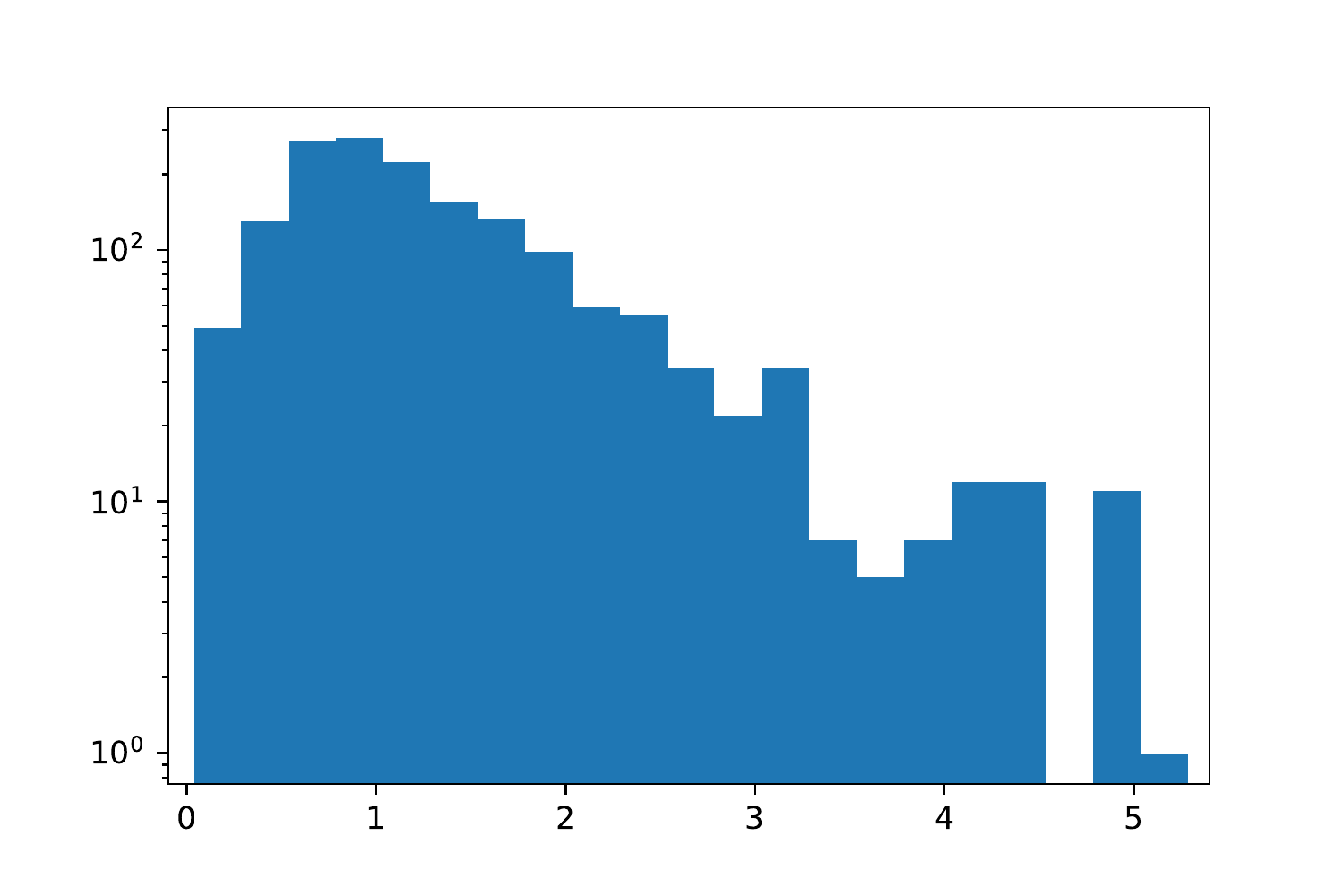}\par
\caption{Redshift distribution of the Risaliti and Lusso 2019 QSO data.}
\label{fig: A Hubble diagram of quasar}
\end{figure}

\section{Method}
\label{sec:methods}
Over the last four decades it has become clear that a quasar's X-ray and UV luminosities are non-linearly correlated \citep{Tan1979, Zam1981, Avni1986, Ste2006, Just2007, Young2010, Lusso2010, Grupe2010, Vagnetti2010}. \cite{Risaliti2015} made use of this correlation to constrain model parameters, as follows. The empirical relation between the quasar's X-ray and UV luminosity is
\begin{equation}
\label{eq:XCDM}
    \log(L_{X}) = \beta + \gamma \log(L_{UV}) ,
\end{equation}
where $\log$ = $\log_{10}$  and $L_{UV}$ and $L_X$ are the QSO UV and X-ray luminosities and $\gamma$ and $\beta$ are  adjustable parameters to be determined from fitting to the measurements. 

What is directly observed are the fluxes and so we need a relation between the UV and X-ray fluxes. Expressing the luminosity in terms of the flux we obtain
\begin{equation}
\label{eq:XCDM}
    \log(F_{X}) = \beta +(\gamma - 1)\log(4\pi) + \gamma \log(F_{UV}) + 2(\gamma - 1)\log(D_L),
\end{equation}
where $F_{UV}$ and $F_X$ are the UV and X-ray fluxes respectively. Here $D_L(z, p)$ is the luminosity distance, which depends on the redshift and the set of cosmological model parameters, $p$, and is given by
\begin{equation}
\label{eq:DM}
  \frac{H_0\sqrt{\left|\Omega_{k0}\right|}D_L(z, p)}{(1+z)} = 
    \begin{cases}
    {\rm sinh}\left[g(z)\right] & \text{if}\ \Omega_{k0} > 0, \\
    \vspace{1mm}
    g(z) & \text{if}\ \Omega_{k0} = 0,\\
    \vspace{1mm}
    {\rm sin}\left[g(z)\right] & \text{if}\ \Omega_{k0} < 0,
    \end{cases}   
\end{equation}
where
\begin{equation}
\label{eq:XCDM}
   g(z) = H_0\sqrt{\left|\Omega_{k0}\right|}\int^z_0 \frac{dz'}{H(z')},
\end{equation}
and the Hubble parameter $H(z)$, which depends on the cosmological parameters, is given in Sec. 2 for each of the six cosmological models we examine in this paper.

\begin{table*}
	\centering
	\caption{Marginalized one-dimensional best-fit parameters and 1$\sigma$ confidence intervals from 2019 and 2015 QSO data for the $H_0 = 68 \pm 2.8$ km s$^{-1}$ Mpc$^{-1}$ prior.}
	\label{tab:BFP}
	\begin{threeparttable}
	\begin{tabular}{lcccccccccc} 
		\hline
		 Data & Model & $\om$ & $\ol$ & $\ok$ & $\omega_{X}$ & $\alpha$ & $H_0$\tnote{a} & $\delta$ & $\beta$ & $\gamma$ \\
		\hline
		2019 QSO data & Flat \lcdm\ & $0.64^{+0.21}_{-0.19}$ & - & - & - & - & $68.00^{+2.80}_{-2.79}$ & $0.23^{+0.004}_{-0.004}$ & $7.58^{+0.33}_{-0.34}$ & $0.62^{+0.01}_{-0.01}$\\
		& Non-flat \lcdm\ & $0.64^{+0.20}_{-0.17}$ & $0.84^{+0.23}_{-0.34}$ & $-0.48^{+0.51}_{-0.43}$ & - & - & $67.95^{+2.79}_{-2.76}$ & $0.23^{+0.004}_{-0.004}$ & $7.91^{+0.41}_{-0.41}$ & $0.61^{+0.01}_{-0.01}$\\
		& Flat XCDM & $0.28^{+0.26}_{-0.14}$ & - & - & $-9.57^{+4.60}_{-6.31}$ & - & $68.02^{+2.76}_{-2.79}$ & $0.23^{+0.004}_{-0.004}$ & $7.78^{+0.31}_{-0.32}$ & $0.62^{+0.01}_{-0.01}$\\
		& Non-flat XCDM & $0.42^{+0.26}_{-0.18}$ & - & $-0.12^{+0.15}_{-0.19}$ & $-5.74^{+2.97}_{-6.43}$ & - & $68.01^{+2.81}_{-2.78}$ & $0.23^{+0.004}_{-0.004}$ & $8.01^{+0.43}_{-0.44}$ & $0.61^{+0.01}_{-0.01}$\\
		&Flat \pcdm\ & $0.61^{+0.20}_{-0.20}$ & - & - & - & $1.30^{+1.11}_{-0.94}$ & $68.01^{+2.81}_{-2.78}$ & $0.23^{+0.004}_{-0.004}$ & $7.59^{+0.33}_{-0.35}$ & $0.62^{+0.01}_{-0.01}$\\
		& Non-flat $\phi$CDM & $0.57^{+0.22}_{-0.20}$ & - & $-0.29^{+0.35}_{-0.27}$ & - & $1.29^{+1.13}_{-0.93}$ & $68.03^{+2.78}_{-2.76}$ & $0.23^{+0.004}_{-0.004}$ & $7.73^{+0.38}_{-0.38}$ & $0.62^{+0.01}_{-0.01}$\\
		\hline
		2015 QSO data\tnote{b}  & Flat \lcdm\ & $0.26^{+0.17}_{-0.11}$ & - & - & - & - & $68.00^{+2.8}_{-2.8}$ & $0.32^{+0.008}_{-0.008}$ & $8.42^{+0.57}_{-0.58}$ & $0.59^{+0.02}_{-0.02}$\\
		& Non-flat \lcdm\ & $0.24^{+0.16}_{-0.10}$ & $0.93^{+0.18}_{-0.39}$ & $-0.17^{+0.49}_{-0.34}$ & - & - & $68.00^{+2.8}_{-2.8}$ & $0.32^{+0.008}_{-0.008}$ & $8.62^{+0.62}_{-0.62}$ & $0.58^{+0.02}_{-0.02}$\\
		& Flat XCDM & $0.25^{+0.16}_{-0.10}$ & - & - & $-2.49^{+1.26}_{-1.59}$ & - & $68.00^{+2.8}_{-2.8}$ & $0.32^{+0.008}_{-0.008}$ & $8.65^{+0.55}_{-0.57}$ & $0.58^{+0.02}_{-0.02}$\\
		& Non-flat XCDM & $0.29^{+0.26}_{-0.14}$ & - & $0.11^{+0.66}_{-0.31}$ & $-1.87^{+1.18}_{-2.05}$ & - & $68.00^{+2.8}_{-2.8}$ & $0.32^{+0.008}_{-0.008}$ & $8.52^{+0.64}_{-0.65}$ & $0.58^{+0.02}_{-0.02}$\\
		&Flat \pcdm\ & $0.26^{+0.18}_{-0.11}$ & - & - & - & $0.54^{+0.43}_{-0.38}$ & $68.00^{+2.8}_{-2.8}$ & $0.32^{+0.008}_{-0.008}$ & $8.42^{+0.57}_{-0.57}$ & $0.59^{+0.02}_{-0.02}$\\
		& Non-flat $\phi$CDM & $0.34^{+0.24}_{-0.16}$ & - & $-0.30^{+0.44}_{-0.61}$ & - & $0.55^{+0.43}_{-0.38}$ & $68.00^{+2.8}_{-2.8}$ & $0.32^{+0.008}_{-0.008}$ & $8.45^{+0.57}_{-0.58}$ & $0.59^{+0.02}_{-0.02}$\\
		\hline
	\end{tabular}
	\begin{tablenotes}
    \item[a]${\rm km}\hspace{1mm}{\rm s}^{-1}{\rm Mpc}^{-1}$.
    \item[b]From \cite{Khadka2019}.
    \end{tablenotes}
    \end{threeparttable}
\end{table*}
\begin{table*}
	\centering
	\caption{Marginalized one-dimensional best-fit parameters and 1$\sigma$ confidence intervals from 2019 and 2015 QSO data for the $H_0 = 73.24 \pm 1.74$ km s$^{-1}$ Mpc$^{-1}$ prior.}
	\label{tab:BFP}
	\begin{threeparttable}
	\begin{tabular}{lcccccccccc} 
		\hline
		Data & Model & $\om$ & $\ol$ & $\ok$ & $\omega_{X}$ & $\alpha$ & $H_0$\tnote{a} & $\delta$ & $\beta$ & $\gamma$ \\
		\hline
		2019 QSO data & Flat \lcdm\ & $0.64^{+0.21}_{-0.19}$ & - & - & - & - & $73.23^{+1.73}_{-1.73}$ & $0.23^{+0.004}_{-0.004}$ & $7.56^{+0.33}_{-0.34}$ & $0.62^{+0.01}_{-0.01}$\\
		& Non-flat \lcdm\ & $0.64^{+0.20}_{-0.17}$ & $0.84^{+0.23}_{-0.34}$ & $-0.48^{+0.51}_{-0.43}$ & - & - & $73.25^{+1.72}_{-1.72}$ & $0.23^{+0.004}_{-0.004}$ & $7.89^{+0.41}_{-0.41}$ & $0.61^{+0.01}_{-0.01}$\\
		& Flat XCDM & $0.28^{+0.26}_{-0.14}$ & - & - & $-9.48^{+4.59}_{-6.40}$ & - & $73.26^{+1.74}_{-1.74}$ & $0.23^{+0.004}_{-0.004}$ & $7.76^{+0.31}_{-0.31}$ & $0.62^{+0.01}_{-0.01}$\\
		& Non-flat XCDM & $0.42^{+0.26}_{-0.19}$ & - & $-0.12^{+0.14}_{-0.19}$ & $-5.74^{+2.93}_{-6.36}$ & - & $73.22^{+1.75}_{-1.72}$ & $0.23^{+0.004}_{-0.004}$ & $8.00^{+0.44}_{-0.45}$ & $0.61^{+0.01}_{-0.01}$\\
		&Flat \pcdm\ & $0.61^{+0.20}_{-0.20}$ & - & - & - & $1.34^{+1.12}_{-0.96}$ & $73.22^{+1.74}_{-1.71}$ & $0.23^{+0.004}_{-0.004}$ & $7.56^{+0.33}_{-0.34}$ & $0.62^{+0.01}_{-0.01}$\\
		& Non-flat $\phi$CDM & $0.56^{+0.22}_{-0.20}$ & - & $-0.34^{+0.37}_{-0.30}$ & - & $1.28^{+1.12}_{-0.91}$ & $73.21^{+1.73}_{-1.71}$ & $0.23^{+0.004}_{-0.004}$ & $7.74^{+0.40}_{-0.40}$ & $0.61^{+0.01}_{-0.01}$\\
		\hline
		2015 QSO data\tnote{b} & Flat \lcdm\ & $0.26^{+0.17}_{-0.11}$ & - & - & - & - & $73.24^{+1.73}_{-1.73}$ & $0.32^{+0.008}_{-0.008}$ & $8.40^{+0.57}_{-0.57}$ & $0.59^{+0.02}_{-0.02}$\\
		& Non-flat \lcdm\ & $0.24^{+0.16}_{-0.10}$ & $0.93^{+0.18}_{-0.39}$ & $-0.17^{+0.49}_{-0.34}$ & - & - & $73.24^{+1.73}_{-1.73}$ & $0.32^{+0.008}_{-0.008}$ & $8.59^{+0.62}_{-0.62}$ & $0.58^{+0.02}_{-0.02}$\\
		& Flat XCDM & $0.25^{+0.16}_{-0.10}$ & - & - & $-2.48^{+1.26}_{-1.59}$ & - & $73.24^{+1.73}_{-1.73}$ & $0.32^{+0.008}_{-0.008}$ & $8.62^{+0.55}_{-0.56}$ & $0.58^{+0.02}_{-0.02}$\\
		& Non-flat XCDM & $0.29^{+0.25}_{-0.14}$ & - & $0.10^{+0.62}_{-0.32}$ & $-1.83^{+1.15}_{-2.02}$ & - & $73.24^{+1.74}_{-1.74}$ & $0.32^{+0.008}_{-0.008}$ & $8.50^{+0.65}_{-0.64}$ & $0.58^{+0.02}_{-0.02}$\\
		&Flat \pcdm\ & $0.24^{+0.19}_{-0.12}$ & - & - & - & $0.55^{+0.43}_{-0.38}$ & $73.23^{+1.73}_{-1.73}$ & $0.32^{+0.008}_{-0.008}$ & $8.40^{+0.57}_{-0.57}$ & $0.59^{+0.02}_{-0.02}$\\
		& Non-flat $\phi$CDM & $0.34^{+0.24}_{-0.17}$ & - & $-0.30^{+0.62}_{-0.44}$ & - & $0.55^{+0.43}_{-0.38}$ & $73.26^{+1.74}_{-1.73}$ & $0.32^{+0.008}_{-0.008}$ & $8.42^{+0.57}_{-0.58}$ & $0.59^{+0.02}_{-0.02}$\\
		\hline
	\end{tabular}
	\begin{tablenotes}
    \item[a]${\rm km}\hspace{1mm}{\rm s}^{-1}{\rm Mpc}^{-1}$.
    \item[b]From \cite{Khadka2019}.
    \end{tablenotes}
    \end{threeparttable}
\end{table*}
To constrain cosmological parameters we compare observed X-ray fluxes to model-predicted X-ray fluxes at the same redshifts. The model-predicted X-ray flux of a QSO depends on the set of cosmological model parameters, the redshift, and the observed UV flux, see eq. (11). We compute the best-fit values and uncertainties of the cosmological parameters of a model by maximizing the likelihood function. The QSO data analysis depends on the $L_{X}-L_{UV}$ relation and this relation has an observed dispersion ($\delta$). So we are required to consider a likelihood function normalization factor which is a function of $\delta$. The QSO data likelihood function $({\rm LF})$ is \citep{Risaliti2015}
\begin{equation}
\label{eq:chi2}
    \ln({\rm LF}) = -\frac{1}{2}\sum^{1598}_{i = 1} \left[\frac{[\log(F^{\rm obs}_{X,i}) - \log(F^{\rm th}_{X,i})]^2}{s^2_i} + \ln(2\pi s^2_i)\right],
\end{equation}
where $\ln$ = $\log_e$ and $s^2_i = \sigma^2_i + \delta^2$, and $\sigma_i$ and $\delta$ are the measurement error on $F^{\rm obs}_{X,i}$ and the global intrinsic dispersion respectively. In eq. (14) $F^{\rm th}_{X,i}$ is the corresponding theoretical model prediction defined by eq. (11), and depends on the observed $F_{UV}$ and $D_L(z_i, p)$. $\delta$ is treated as a free parameter to be determined by the data, along with the other two free parameters,  $\gamma$ and $\beta$, that characterise the $L_X$ - $L_{UV}$ relation in eq. (10). In \cite{Risaliti2019}, also see \cite{Lusso2019}, $\gamma$ is not a free parameter, $\beta$ is determined by calibrating quasar distance modulus using JLA supernovae data over the common redshift range $z < 1.4$, and $\delta$ is a free parameter, whereas in \cite{wei2020} $\beta$ is determined by calibrating quasar distance modulus using Hubble parameter measurements, and $\gamma$ and $\delta$ are free parameters. We instead follow \cite{Khadka2019} and treat $\beta$, $\gamma$, and $\delta$ as free parameters to be determined, along with the cosmological parameters, from the QSO data, in each cosmological model. As a consequence, our QSO constraints are QSO-only constraints (they do not make use of the supernovae or $H(z)$ data),\footnote{As discussed below, we do use two different $H_0$ priors for analysing the QSO data, however the derived QSO constraints on parameters, excluding that on $H_0$, are almost insensitive to the choice of $H_0$ prior.} which makes them a little less constraining than the \cite{Risaliti2019} results, but allows us to compare QSO-only constraints to those from other data.

Our determination of the $H(z)$ and BAO data constraints uses the procedure outlined in Sec. 4 of \cite{Khadka2019}.

For every parameter except $H_0$, we use top-hat priors, that are non-zero over the ranges $0 \leq \om \leq 1$, $0 \leq \ol \leq 1.3$, $-0.7 \leq k \leq 0.7$, $-20 \leq \omega_X \leq 5$, $0 \leq \alpha \leq 3$ , $-10 \leq \ln{\delta} \leq 10$, $0 \leq \beta \leq 11$, and $-2 \leq \gamma \leq 2$. Here $k$ = $-\Omega_{k0} a^2_0$ where $a_0$ is the current value of the scale factor. For $H_0$ we consider two different Gaussian priors, $H_0 = 68 \pm 2.8$ km s$^{-1}$ Mpc$^{-1}$, fron a median statistics analysis of a large compilation of $H_0$ measurements \citep{chen3},\footnote{This value is very consistent with those from earlier median statistics analyses \citep{Gott2001, chen1}, and with many recent measurements of $H_0$ \citep{chen5, DESb, Yu2018, Gomez2018, Haridasu2018, Plank2018, Zhang2018a, D, Martinelli2019, Cuceu2019, Zeng2019, schon2019, LinI2019, Zhang2019}. While the \cite{Plank2018} cosmic microwave background (CMB) anisotropy data more tightly constrains $H_0$, these constraints depend on the cosmological model used to analyze the CMB data.} and $H_0 = 73.24 \pm 1.74$ km s$^{-1}$ Mpc$^{-1}$, from a recent local expansion rate measurement \citep{Riess2016}.\footnote{Other local expansion rate determinations result in somewhat  lower $H_0$ values with somewhat larger error bars \citep{Rigault2015, Zhang2017, Dhaw, Fernandez2018, Freedman2019, Freedman2020, Rameez2019}.}

The likelihood analysis is done using the Markov chain Monte Carlo (MCMC) method implemented in the emcee package \citep{Foreman2013} in Python 3.7.

For the QSO data we use the maximum likelihood value $\rm LF_{\rm max}$ to compute the minimum $\chi^2_{\rm min, QSO}$ = $-2\ln{(\rm LF_{\rm max, \rm QSO})} - \sum^{1598}_{i = 1}\ln(2\pi (\sigma^2_{i, \rm QSO} + \delta^2_{\rm bestfit}))$.\footnote{In \cite{Khadka2019}, the  $\chi^2_{\rm min}$ for the QSO data was incorrectly computed  using the conventional minimum $-2\ln{(\rm LF_{\rm max})}$. This resulted in an incorrect, low, reduced $\chi^2_{\rm min}$ for the 2015 QSO data, < 0.6, see Tables 1 and 2 of \cite{Khadka2019}. Including the normalization factor in the computation of $\chi^2_{\rm min}$ for the 2015 QSO data, the reduced $\chi^2_{\rm min}$ are very close to unity in all models.} The second term in the expression for $\chi^2_{\rm min, \rm QSO}$ is a consequence of the normalization factor in the QSO likelihood function, see eq. (14). The $\chi^2_{\rm min}$ for the QSO + BAO + $H(z)$ data set also accounts for the QSO normalization factor, while in the case of the $H(z)$ + BAO data set we compute the conventional minimum $\chi^2_{\rm min, \rm H(z) + BAO}$ = $-2\ln{(\rm LF_{\rm max, \rm H(z) + BAO})}$. In addition to $\chi^2_{\rm min}$ we compute the Akaike Information Criterion
\begin{equation}
\label{eq:AIC}
    AIC = \chi^2_{\rm min} + 2d ,
\end{equation}
as well as the Bayes Information Criterion
\begin{equation}
\label{eq:AIC}
    BIC = \chi^2_{\rm min} + d\ln{N},
\end{equation}
where $d$ is the number of free model parameters, $N$ is the number of data points, and we define the degrees of freedom dof = $N - d$. The $AIC$ and $BIC$ penalize models that have more free parameters.
\begin{table*}
	\centering
	\caption{Unmarginalized best-fit parameters for the $H_0 = 68 \pm 2.8$ km s$^{-1}$ Mpc$^{-1}$ prior.}
	\label{tab:BFP}
	\begin{threeparttable}
	\begin{tabular}{lcccccccccccccc} 
		\hline
		Model & Data set & $\om$ & $\ol$ & $\ok$ & $\omega_{X}$ & $\alpha$ & $H_0$\tnote{a} & $\delta$ & $\beta$ & $\gamma$ & $\chi^2_{\rm min}$ & dof & $AIC$ & $BIC$\\
		\hline
		Flat \lcdm\ &  $H(z)$ + BAO\tnote{b} & 0.29 & 0.71 & - & - & - & 67.56 & - & - & - & 32.47 & 40 & 36.47 & 39.95\\
		 & QSO & 0.60 & 0.40 & - & - & - & 68.00 & 0.23 & 7.57 & 0.62 & 1606.99 & 1593 & 1616.99 & 1643.87\\
		 & QSO + $H(z)$ + BAO & 0.30 & 0.70 & - & - & - & 68.03 & 0.23 & 7.12 & 0.64 & 1630.00 &  1635 & 1640.00 & 1667.01\\
		\hline
		Non-flat \lcdm\ &  $H(z)$ + BAO\tnote{b} & 0.30 & 0.70 & $0.00$ & - & - & 68.23 & - & - & - & 27.05 & 39 & 33.05 & 38.26\\
		 & QSO & 0.56 & 0.98 & $-0.54$ & - & - & 68.00 & 0.23 & 7.93 & 0.61 & 1604.37 & 1592 & 1616.37 & 1648.63\\
		 & QSO + $H(z)$ + BAO & 0.30 & 0.71 & $-0.01$ & - & - & 68.77 & 0.23 & 7.11 & 0.64 & 1630.00 & 1634 & 1642.00 & 1674.41\\
		\hline
		Flat XCDM &  $H(z)$ + BAO\tnote{b} & 0.30 & 0.70 & - & $-0.96$ & - & 67.24 & - & - & - & 27.29 & 39 & 33.29 & 38.50\\
		 & QSO & 0.20 & 0.80 & - & $-7.08$ & - & 68.00 & 0.23 & 7.66 & 0.62 & 1603.01 & 1592 & 1615.01 & 1647.27\\
		 & QSO + $H(z)$ + BAO & 0.30 & 0.70 & - & $-0.96$ & - & 67.30 & 0.23 & 7.13 & 0.64 & 1629.76 & 1634 & 1641.76 & 1674.17\\
		 \hline
		Non-flat XCDM & $H(z)$ + BAO\tnote{b} & 0.32 & - & $-0.23$ & $-0.74$ & - & 67.42 & - & - & - & 24.91 & 38 & 32.91 & 39.86\\
		 & QSO & 0.29 & - & $-0.15$ & $-4.87$ & - & 68.00 & 0.23 & 8.10 & 0.61 & 1604.29 & 1591 & 1618.29 & 1655.93\\
		 & QSO + $H(z)$ + BAO & 0.33 & - & $-0.40$ & $-0.66$ & - & 67.43 & 0.23 & 7.54 & 0.62 & 1628.82 & 1633 & 1642.82 & 1680.64\\
		\hline
		Flat \pcdm\ & $H(z)$ + BAO\tnote{b} & 0.32 & - & - & - & 0.10 & 67.23 & - & - & - & 27.42 & 39 & 33.42 & 38.63\\
		 & QSO & 0.82 & - & - & - & 2.03 & 68.19 & 0.23 & 7.77 & 0.61 & 1589.32 & 1592 & 1601.32 & 1633.58\\
		 & QSO + $H(z)$ + BAO & 0.30 & - & - & - & 0.09 & 67.62 & 0.23 & 7.21 & 0.64 & 1633.40 & 1634 & 1645.40 & 1677.81\\
		\hline
		Non-flat $\phi$CDM & $H(z)$ + BAO\tnote{b} & 0.33 & - & $-0.20$ & - & 1.20 & 65.86 & - & - & - & 25.04 & 38 & 33.04 & 39.99\\
		 & QSO & 0.56 & - & $-0.55$ & - & 0.08 & 67.63 & 0.23 & 7.99 & 0.61 & 1626.71 & 1591 & 1640.71 & 1678.35\\
		 & QSO + $H(z)$ + BAO & 0.32 & - & $-0.41$ & - & 1.51 & 67.81 & 0.23 & 7.54 & 0.62 & 1624.67 & 1633 & 1639.67 & 1676.49\\
		 \hline
	\end{tabular}
	\begin{tablenotes}
    \item[a]${\rm km}\hspace{1mm}{\rm s}^{-1}{\rm Mpc}^{-1}$.
    \item[b]From \cite{Khadka2019}.
    \end{tablenotes}
    \end{threeparttable}
\end{table*}

\begin{table*}
	\centering
	\caption{Unmarginalized best-fit parameters for the $H_0 = 73.24 \pm 1.74$ km s$^{-1}$ Mpc$^{-1}$ prior.}
	\label{tab:BFP}
	\begin{threeparttable}
	\begin{tabular}{lcccccccccccccc} 
		\hline
		Model & Data set & $\om$ & $\ol$ & $\ok$ & $\omega_{X}$ & $\alpha$ & $H_0$\tnote{a} & $\delta$  & $\beta$ & $\gamma$ & $\chi^2_{\rm min}$ & dof & $AIC$ & $BIC$\\
		\hline
		Flat \lcdm\ &  $H(z)$ + BAO\tnote{b} & 0.30 & 0.70 & - & - & - & 69.11 & - & - & - & 33.76 & 40 & 38.76 & 41.24\\
		 & QSO & 0.60 & 0.40 & - & - & - & 73.24 & 0.23 & 7.54 & 0.62 & 1606.03 & 1593 & 1616.03 & 1642.91\\
		 & QSO + $H(z)$ + BAO & 0.31 & 0.69 & - & - & - & 69.15 & 0.23 & 7.12 & 0.64 & 1636.26 & 1635 & 1646.26 & 1673.27\\
		\hline
		Non-flat \lcdm\ &  $H(z)$ + BAO\tnote{b} & 0.30 & 0.78 & $-0.08$ & - & - & 71.56 & - & - & - & 28.80 & 39 & 34.80 & 40.01\\
		 & QSO & 0.56 & 0.98 & $-0.54$ & - & - & 73.24 & 0.23 & 7.91 & 0.61 & 1604.37 & 1592 & 1616.37 & 1648.78\\
		 & QSO + $H(z)$ + BAO & 0.31 & 0.79 & $-0.1$ & - & - & 71.85 & 0.23 & 7.16 & 0.64 & 1631.48 & 1634 & 1643.48 & 1675.89\\
		\hline
		Flat XCDM &  $H(z)$ + BAO\tnote{b} & 0.29 & 0.71 & - & $-1.14$ & - & 71.27 & - & - & - & 30.68 & 39 & 36.68 & 41.89\\
		 & QSO & 0.20 & 0.80 & - & $-7.08$ & - & 73.24 & 0.23 & 7.64 & 0.62 & 1603.01 & 1592 & 1615.01 & 1647.27\\
		 & QSO + $H(z)$ + BAO & 0.30 & 0.70 & - & $-1.14$ & - & 71.32 & 0.23 & 7.13 & 0.64 & 1633.16 & 1634 & 1645.16 & 1677.57\\
		 \hline
		Non-flat XCDM & $H(z)$ + BAO\tnote{b} & 0.32 & - & $-0.21$ & $-0.85$ & - & 71.22 & - & - & - & 28.17 & 38 & 36.17 & 43.12\\
		 & QSO & 0.29 & - & $-0.15$ & $-4.87$ & - & 73.24 & 0.23 & 8.08 & 0.61 & 1604.29 & 1591 & 1618.29 & 1655.93\\
		 & QSO + $H(z)$ + BAO & 0.33 & - & $-0.38$ & $-0.74$ & - & 71.11 & 0.23 & 7.47 & 0.63 & 1632.09 & 1633 & 1646.09 & 1683.91\\
		\hline
		Flat \pcdm\ & $H(z)$ + BAO\tnote{b} & 0.33 & - & - & - & 0.09 & 69.31 & - & - & - & 33.36 & 39 & 39.36 & 44.57\\
		 & QSO & 0.61 & - & - & - & 0.26 & 73.11 & 0.23 & 7.53 & 0.62 & 1601.22 & 1592 & 1613.22 & 1645.48\\
		 & QSO + $H(z)$ + BAO & 0.31 & - & - & - & 0.003 & 69.40 & 0.23 & 7.17 & 0.63 & 1636.87 & 1634 & 1638.87 & 1671.28\\
		\hline
		Non-flat $\phi$CDM & $H(z)$ + BAO\tnote{b} & 0.32 & - & $-0.22$ & - & 1.14 & 69.23 & - & - & - & 27.62 & 38 & 35.62 & 42.57\\
		 & QSO & 0.49 & - & $-0.53$ & - & 0.01 & 72.98 & 0.23 & 7.78 & 0.61 & 1606.10 & 1591 & 1620.10 & 1657.74\\
		 & QSO + $H(z)$ + BAO & 0.32 & - & $-0.39$ & - & 1.09 & 71.22 & 0.23 & 7.47 & 0.63 & 1640.19 & 1633 & 1654.19 & 1692.01\\
		 \hline
	\end{tabular}
	\begin{tablenotes}
    \item[a]${\rm km}\hspace{1mm}{\rm s}^{-1}{\rm Mpc}^{-1}$.
    \item[b]From \cite{Khadka2019}.
    \end{tablenotes}
    \end{threeparttable}
\end{table*}
\begin{table*}
	\centering
	\caption{Marginalized one-dimensional best-fit parameters with 1$\sigma$ confidence intervals for all models using BAO and $H(z)$ data \citep[from][]{Khadka2019}.}
	\label{tab:BFP}
	\begin{threeparttable}
	\begin{tabular}{lcccccccccc} 
		\hline
		$H_0$\tnote{a}\hspace{3mm}prior & Model & $\om$ & $\ol$ & $\ok$ & $\omega_{X}$ & $\alpha$ & $H_0$\tnote{a}\\
		\hline
		$H_0 = 68 \pm 2.8$ & Flat \lcdm\ & $0.29^{+0.01}_{-0.01}$ & - & - & - & - & $67.58^{+0.85}_{-0.85}$ \\
		& Non-flat \lcdm\ & $0.30^{+0.01}_{-0.01}$ & $0.70^{+0.05}_{-0.06}$ & $0.00^{+0.06}_{-0.07}$ & - & - & $68.17^{+1.80}_{-1.79}$\\
		& Flat XCDM & $0.30^{+0.02}_{-0.02}$ & - & - & $-0.97^{+0.09}_{-0.09}$ & - & $67.39^{+1.87}_{-1.84}$\\
		& Non-flat XCDM & $0.32^{+0.02}_{-0.02}$ & - & $-0.18^{+0.17}_{-0.21}$ & $-0.77^{+0.11}_{-0.17}$ & - & $67.42^{+1.84}_{-1.80}$\\
		&Flat \pcdm\ & $0.31^{+0.01}_{-0.01}$ & - & - & - & $0.20^{+0.21}_{-0.13}$ & $66.57^{+1.31}_{-1.46}$\\
		& Non-flat $\phi$CDM & $0.31^{+0.01}_{-0.01}$ & - & $-0.20^{+0.13}_{-0.17}$ & - & $0.86^{+0.55}_{-0.49}$ & $67.69^{+1.75}_{-1.74}$\\
		\hline
		$H_0 = 73.24 \pm 1.74$ & Flat \lcdm\ & $0.31^{+0.01}_{-0.01}$ & - & - & - & - & $69.12^{+0.81}_{-0.80}$\\
		& Non-flat \lcdm\ & $0.30^{+0.01}_{-0.01}$ & $0.78^{+0.04}_{-0.04}$ & $-0.08^{+0.05}_{-0.05}$ & - & - & $71.51^{+1.41}_{-1.40}$\\
		& Flat XCDM & $0.29^{+0.02}_{-0.01}$ & - & - & $-1.14^{+0.08}_{-0.08}$ & - & $71.32^{+1.49}_{-1.48}$\\
		& Non-flat XCDM & $0.32^{+0.02}_{-0.02}$ & - & $-0.17^{+0.16}_{-0.19}$ & $-0.88^{+0.14}_{-0.21}$ & - & $71.23^{+1.46}_{-1.46}$\\
		&Flat \pcdm\ & $0.31^{+0.01}_{-0.01}$ & - & - & - & $0.07^{+0.09}_{-0.04}$ & $68.91^{+0.98}_{-1.00}$\\
		& Non-flat $\phi$CDM & $0.32^{+0.01}_{-0.01}$ & - & $-0.25^{+0.12}_{-0.16}$ & - & $0.68^{+0.53}_{-0.46}$ & $71.14^{+1.39}_{-1.38}$\\
		\hline
	\end{tabular}
	\begin{tablenotes}
    \item[a]${\rm km}\hspace{1mm}{\rm s}^{-1}{\rm Mpc}^{-1}$.
    \end{tablenotes}
    \end{threeparttable}
\end{table*}
\begin{table*}
	\centering
	\caption{Marginalized one-dimensional best-fit parameters with 1$\sigma$ confidence intervals for all models using  QSO+$H(z)$+BAO data.}
	\label{tab:BFP}
	\begin{threeparttable}
	\begin{tabular}{lcccccccccc} 
		\hline
		$H_0$\tnote{a}\hspace{3mm}prior & Model & $\om$ & $\ol$ & $\ok$ & $\omega_{X}$ & $\alpha$ & $H_0$\tnote{a} & $\delta$ & $\beta$ & $\gamma$\\
		\hline
		$H_0 = 68 \pm 2.8$ & Flat \lcdm\ & $0.30^{+0.01}_{-0.01}$ & $0.70^{+0.01}_{-0.01}$ & - & - & - & $68.04^{+0.84}_{-0.84}$ & $0.23^{+0.004}_{-0.004}$ & $7.11^{+0.27}_{-0.27}$ & $0.64^{+0.009}_{-0.009}$\\
		& Non-flat \lcdm\ & $0.30^{+0.01}_{-0.01}$ & $0.71^{+0.05}_{-0.06}$ & $-0.01^{+0.06}_{-0.07}$ & - & - & $68.70^{+1.78}_{-1.79}$ & $0.23^{+0.004}_{-0.004}$ & $7.11^{+0.27}_{-0.27}$ & $0.64^{+0.009}_{-0.009}$\\
		& Flat XCDM & $0.30^{+0.02}_{-0.02}$ & - & - & $-0.96^{+0.09}_{-0.09}$ & - & $67.41^{+1.88}_{-1.83}$ & $0.23^{+0.004}_{-0.004}$ & $7.12^{+0.27}_{-0.27}$ & $0.64^{+0.009}_{-0.009}$\\
		& Non-flat XCDM & $0.33^{+0.02}_{-0.02}$ & - & $-0.34^{+0.18}_{-0.18}$ & $-0.69^{+0.07}_{-0.11}$ & - & $67.48^{+1.81}_{-1.77}$ & $0.23^{+0.004}_{-0.004}$ & $7.47^{+0.33}_{-0.33}$ & $0.63^{+0.01}_{-0.01}$\\
		&Flat \pcdm\ & $0.31^{+0.01}_{-0.01}$ & - & - & - & $0.20^{+0.21}_{-0.14}$ & $66.76^{+1.36}_{-1.49}$ & $0.23^{+0.004}_{-0.004}$ & $7.16^{+0.27}_{-0.27}$ & $0.64^{+0.009}_{-0.009}$\\
		& Non-flat $\phi$CDM & $0.32^{+0.01}_{-0.01}$ & - & $-0.32^{+0.16}_{-0.16}$ & - & $1.21^{+0.47}_{-0.53}$ & $67.90^{+1.72}_{-1.73}$ & $0.23^{+0.004}_{-0.004}$ & $7.47^{+0.33}_{-0.32}$ & $0.63^{+0.01}_{-0.01}$\\
		\hline
		$H_0 = 73.24 \pm 1.74$ & Flat \lcdm\ & $0.31^{+0.01}_{-0.01}$ & $0.69^{+0.01}_{-0.01}$ & - & - & - & $69.16^{+0.81}_{-0.81}$ & $0.23^{+0.004}_{-0.004}$ & $7.12^{+0.27}_{-0.27}$ & $0.64^{+0.009}_{-0.009}$\\
		& Non-flat \lcdm\ & $0.31^{+0.01}_{-0.01}$ & $0.78^{+0.04}_{-0.04}$ & $-0.09^{+0.05}_{-0.05}$ & - & - & $71.79^{+1.40}_{-1.39}$ & $0.23^{+0.004}_{-0.004}$ & $7.16^{+0.27}_{-0.27}$ & $0.64^{+0.009}_{-0.009}$\\
		& Flat XCDM & $0.30^{+0.02}_{-0.01}$ & - & - & $-1.14^{+0.08}_{-0.08}$ & - & $71.38^{+1.51}_{-1.50}$ & $0.23^{+0.004}_{-0.004}$ & $7.09^{+0.27}_{-0.27}$ & $0.64^{+0.009}_{-0.009}$\\
		& Non-flat XCDM & $0.33^{+0.02}_{-0.02}$ & - & $-0.31^{+0.17}_{-0.18}$ & $-0.77^{+0.09}_{-0.15}$ & - & $71.17^{+1.45}_{-1.43}$ & $0.23^{+0.004}_{-0.004}$ & $7.41^{+0.34}_{-0.33}$ & $0.63^{+0.01}_{-0.01}$\\
		&Flat \pcdm\ & $0.31^{+0.01}_{-0.01}$ & - & - & - & $0.06^{+0.09}_{-0.05}$ & $69.09^{+1.01}_{-1.02}$ & $0.23^{+0.004}_{-0.004}$ & $7.15^{+0.27}_{-0.27}$ & $0.64^{+0.009}_{-0.009}$\\
		& Non-flat $\phi$CDM & $0.32^{+0.01}_{-0.01}$ & - & $-0.35^{+0.15}_{-0.15}$ & - & $0.98^{+0.44}_{-0.50}$ & $71.24^{+1.40}_{-1.39}$ & $0.23^{+0.004}_{-0.004}$ & $7.47^{+0.33}_{-0.32}$ & $0.63^{+0.01}_{-0.01}$\\
		\hline
	\end{tabular}
	\begin{tablenotes}
    \item[a]${\rm km}\hspace{1mm}{\rm s}^{-1}{\rm Mpc}^{-1}$.
    \end{tablenotes}
    \end{threeparttable}
\end{table*}
\begin{figure}
    \includegraphics[width=\linewidth,right]{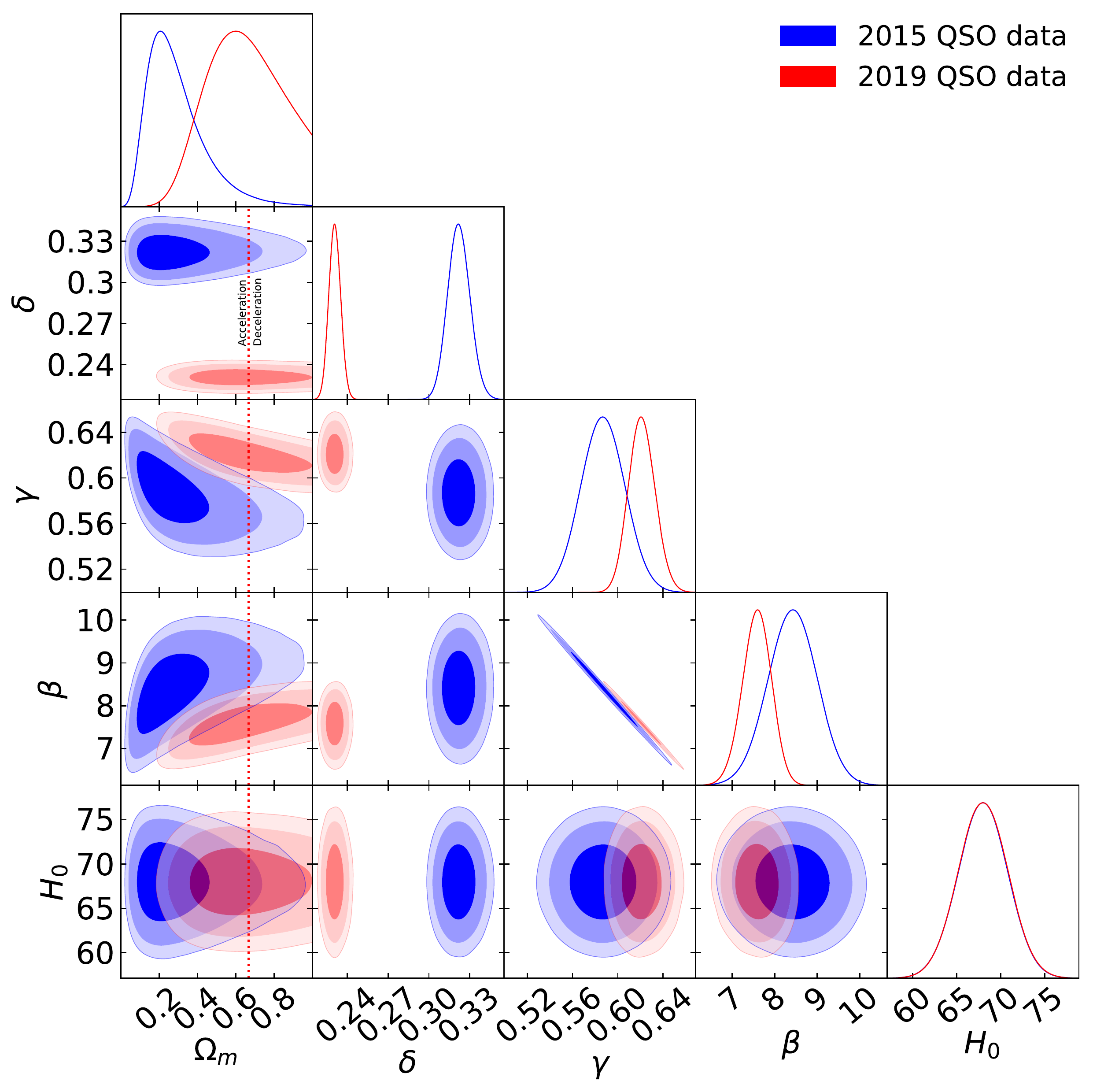}\par
\caption{Flat $\Lambda$CDM model constraints from the 2015 QSO data (blue) and the 2019 QSO data (red) for the $H_0 = 68 \pm 2.8$ ${\rm km}\hspace{1mm}{\rm s}^{-1}{\rm Mpc}^{-1}$ prior. Shown are 1, 2, and 3$\sigma$ confidence contours and one-dimensional likelihoods for all free parameters. The red dotted vertical straight lines in the left column of panels are zero acceleration lines, with the current cosmological expansion accelerating to the left of the line where $\om < 0.67$.}
\label{fig: comparison plot}
\end{figure}
\begin{figure}
    \includegraphics[width=\linewidth]{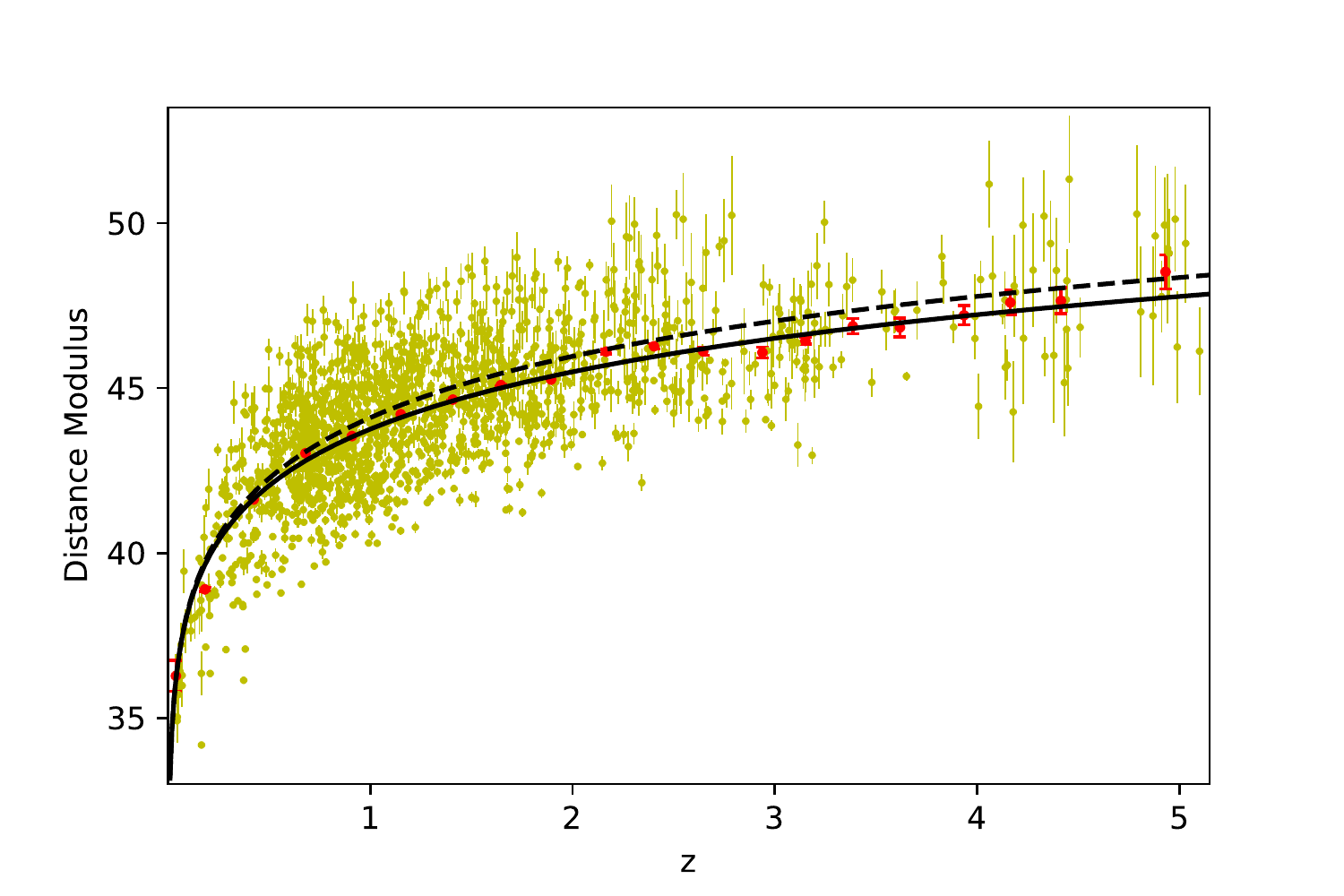}\par
\caption{Hubble diagram of quasars using the flat $\Lambda$CDM model. Black solid line is the best-fit flat $\Lambda$CDM \textbf{model with $\om$ = 0.60} from the 2019 QSO data. Red points are the means and uncertainties on the mean of the distance modulus in narrow redshift bins for the quasar data. These averages do not play a role in the statistical analysis and are shown only for visualization purposes. The black dashed line shows a flat $\Lambda$CDM model with $\om$ = 0.30.}
\label{fig: A Hubble diagram of quasar}
\end{figure}
\section{Results}
\label{sec:Results}
\subsection{Comparison of 2015 and 2019 QSO data constraints}
\label{sec:QSO comparison}
QSO constraints obtained from the 2015 QSO data \citep{Khadka2019} and the 2019 QSO data are largely consistent with each other but there are some differences, including some significant ones. Tables 1 and 2 list best-fit parameter values and 1$\sigma$ error bars determined from the 2019 and 2015 QSO data, for the two different $H_0$ priors. Best-fit values of parameters related to the $L_X-L_{UV}$ relation ($\delta$, $\beta$, and $\gamma$) have changed in comparison to those obtained from the 2015 QSO data. $\beta$ and $\gamma$ are the intercept and slope of the $L_X-L_{UV}$ relation and their values do not tell how well this relation fits the data; the value of the intrinsic dispersion ($\delta$) quantifies how well the $L_X-L_{UV}$ relation fits the data. The intrinsic dispersion of the $L_X-L_{UV}$ relation obtained from the 2015 QSO data and 2019 QSO data are $0.32 \pm 0.008$ and $0.23 \pm 0.004$ respectively, independent of $H_0$ prior and cosmological model. This shows that the 2019 QSO data are described by a tighter $L_X-L_{UV}$ relation than that for the 2015 data. This could be the result of the modified sample filtering process adopted in \cite{Risaliti2019}.

In the case of cosmological parameters, the best-fit values of the equation of state parameter ($\omega_X$) in the flat and non-flat XCDM parametrization obtained from the 2019 QSO data are significantly more negative than those obtained from the 2015 QSO data. From Tables 1 and 2, the 2019 QSO data indicate that the dark energy density in the XCDM parametrization increases with time. Another notable difference between the 2015 QSO data and the 2019 QSO data is that the 2015 QSO data favor a smaller value of the matter density parameter ($\Omega_{m0} \sim 0.3$), consistent with values obtained from other cosmological probes, while the 2019 QSO data favor a larger value of the matter density parameter ($\Omega_{m0} > 0.42$), with the exception of the flat XCDM case where the 2019 data also favor $\om \sim 0.30$. This can be seen in Tables 1 and 2 and  Fig.\ 2 which shows the constraints for the flat $\Lambda$CDM model with the $H_0 = 68 \pm 2.8$ ${\rm km}\hspace{1mm}{\rm s}^{-1}{\rm Mpc}^{-1}$ prior. We note that both high redshift cosmic microwave background anisotropy data \citep{Plank2018} and low redshift, $z < 2.5$, data \citep{chen03, Park2018d} are both consistent with $\om \sim 0.30$ in a variety of different cosmological models, so it is somewhat surprising that the 2019 QSO data at $z \sim 2 - 5$ largely  favor $\om \sim 0.4 - 0.6$.\footnote{We note that our result differs significantly from \cite{Melia2019}, Table 1, who finds $\om = 0.31 \pm 0.05$ in the flat $\Lambda$CDM model from the 2019 QSO data (which is identical to the \cite{Risaliti2019} value of $\om = 0.30 \pm 0.05$ determined from the $z < 1.4$ 2019 QSO data with the JLA supernovae data). The more approximate analyses of \cite{yang2019} and \cite{Velten2020} find larger $\om$ values, as does the analyses of \cite{wei2020} in which they use $H(z)$ data to calibrate the 2019 QSO data. From their more approximate analyses \cite{Velten2020} conclude that the 2019 QSO data are incompatible with a currently accelerating cosmological expansion, Our more accurate analyses shows that while part of the probability lies in the non-accelerating region of cosmological parameter space, in most models we study here a significant part of the probability lies in the accelerating part of cosmological parameter space, see Fig.\ 2 for the flat $\Lambda$CDM case and later figures for other models, and so it is incorrect to claim that the 2019 QSO data are incompatible with currently accelerated cosmological expansion.} It is probably more likely that this larger $\om$ is a reflection of something related to the 2019 QSO data than an indication of the invalidity of the $\Lambda$CDM scenario. A larger value of the matter density parameter gives a lower distance modulus for an astrophysical object at any redshift. So the Hubble diagram of quasars obtained from the 2019 QSO data lies below the Hubble diagram obtained from the concordance model (flat $\Lambda$CDM) with non-relativistic matter density parameter $\Omega_{m0} = 0.30$ and the difference increases with increasing redshift. This can be seen in Fig.\ 3. Qualitatively, Fig.\ 3 shows that QSO data at $z \lesssim 2 $ are consistent with an $\Omega_{m0} = 0.30$ model while the QSO data at $z \gtrsim 2$ favor the $\Omega_{m0} = 0.60$ model. This is qualitatively consistent with the findings of \cite{Risaliti2019}.

\subsection{2019 QSO constraints}
\label{sec:QSO}
The observed correlation between a quasar's X-ray and UV measurements, eq. (10), provides an opportunity to use QSO measurements to constrain cosmological parameters. The global intrinsic dispersion ($\delta$) obtained here is smaller than that of \cite{Khadka2019} for the 2015 QSO data but it still is large and so parameter determination performed using these measurements is not as precise as that done using other data such as BAO or $H(z)$ measurements. But the main advantage of using the QSO data is that it covers a very large range of redshift, part of which is not well probed by other data, so it provides the opportunity of testing cosmological models in a new, higher, redshift range, and it is likely that future, improved, QSO data will provide significant and interesting constraints on cosmological parameters.

The QSO measurements determined cosmological model parameter results are listed in Tables 1--4. The unmarginalized best-fit parameters are listed in the Tables 3 and 4 for the $H_0 = 68 \pm 2.8$ ${\rm km}\hspace{1mm}{\rm s}^{-1}{\rm Mpc}^{-1}$ and $73.24 \pm 1.74$ ${\rm km}\hspace{1mm}{\rm s}^{-1}{\rm Mpc}^{-1}$ priors respectively. The two-dimensional confidence contours and the one-dimensional likelihoods are shown in grey in the left panels of Figs.\ 4--15. The cosmological parameter constraints are almost insensitive to the $H_0$ prior used. For the QSO data, from Tables 1 and 2, the non-relativistic matter density parameter is measured to lie in the range $\om$ = $0.28^{+0.26}_{-0.14}$ to $0.64^{+0.21}_{-0.19}$ ($0.42^{+0.26}_{-0.18}$ to $0.64^{+0.20}_{-0.17}$)  for flat (non-flat) models and the $H_0 = 68 \pm 2.8$ ${\rm km}\hspace{1mm}{\rm s}^{-1}{\rm Mpc}^{-1}$ prior and to lie in the range $\om$ = $0.28^{+0.26}_{-0.14}$ to $0.64^{+0.21}_{-0.19}$ ($0.42^{+0.26}_{-0.19}$ to $0.64^{+0.20}_{-0.17}$)
for flat (non-flat) models and the $H_0 = 73.24 \pm 1.74$ ${\rm km}\hspace{1mm}{\rm s}^{-1}{\rm Mpc}^{-1}$ prior. While the errors are large, the values of $\om$ obtained from the 2019 QSO data in most models are larger than those obtained from other cosmological probes.

From Tables 1 and 2, for the non-flat $\Lambda$CDM model the curvature energy density parameter is measured to be $\ok$ = $-0.48^{+0.51}_{-0.43}$ ($-0.48^{+0.51}_{-0.43}$) for the $H_0 = 68 \pm 2.8$ ${\rm km}\hspace{1mm}{\rm s}^{-1}{\rm Mpc}^{-1}$($73.24 \pm 1.74$ ${\rm km}\hspace{1mm}{\rm s}^{-1}{\rm Mpc}^{-1}$) prior. For the non-flat XCDM model we find $\ok$ = $-0.12^{+0.15}_{-0.19}$ ($-0.12^{+0.14}_{-0.19}$) for the $H_0 = 68 \pm 2.8$ ${\rm km}\hspace{1mm}{\rm s}^{-1}{\rm Mpc}^{-1}$($73.24 \pm 1.74$ ${\rm km}\hspace{1mm}{\rm s}^{-1}{\rm Mpc}^{-1}$) prior. For the non-flat $\phi$CDM model we find $\ok$ = $-0.29^{+0.35}_{-0.27}$ ($-0.34^{+0.37}_{-0.30}$) for the $H_0 = 68 \pm 2.8$ ${\rm km}\hspace{1mm}{\rm s}^{-1}{\rm Mpc}^{-1}$($73.24 \pm 1.74$ ${\rm km}\hspace{1mm}{\rm s}^{-1}{\rm Mpc}^{-1}$) prior. In all models closed spatial hypersurfaces are weakly favored.

From Tables 1 and 2, for the flat (non-flat) $\Lambda$CDM model the dark energy density parameter is $\ol$ = $0.36^{+0.19}_{-0.21}$ ($0.84^{+0.23}_{-0.34}$) for both $H_0 = 68 \pm 2.8$ ${\rm km}\hspace{1mm}{\rm s}^{-1}{\rm Mpc}^{-1}$ and $73.24 \pm 1.74$ ${\rm km}\hspace{1mm}{\rm s}^{-1}{\rm Mpc}^{-1}$ priors.

The equation of state parameter for the flat (non-flat) XCDM model is $\omega_{X}$ = $-9.57^{+4.60}_{-6.31}$ ($-5.74^{+2.97}_{-6.43}$) for the $H_0 = 68 \pm 2.8$ ${\rm km}\hspace{1mm}{\rm s}^{-1}{\rm Mpc}^{-1}$ prior and $-9.48^{+4.59}_{-6.40}$ ($-5.74^{+2.93}_{-6.36}$) for the $73.24 \pm 1.74$ ${\rm km}\hspace{1mm}{\rm s}^{-1}{\rm Mpc}^{-1}$ prior. For both priors $\omega_{X}$ is very low in comparison to the 2015 QSO data values obtained in \cite{Khadka2019}. In the XCDM parametrization the 2019 QSO data favors dark energy density that increases with time. The $\alpha$ parameter in the flat (non-flat) $\phi$CDM model is $\alpha$ = $1.30^{+1.11}_{-0.94}$ ($1.29^{+1.13}_{-0.93}$) for the $H_0 = 68 \pm 2.8$ ${\rm km}\hspace{1mm}{\rm s}^{-1}{\rm Mpc}^{-1}$ prior and $1.34^{+1.12}_{-0.96}$ ($1.28^{+1.12}_{-0.91}$) for the $73.24 \pm 1.74$ ${\rm km}\hspace{1mm}{\rm s}^{-1}{\rm Mpc}^{-1}$ prior. In both models dynamical dark energy is favored.

From the $\chi^2_{\rm min}$, AIC, and BIC values for the QSO data listed in Tables 3 and 4, independent of $H_0$ prior, the flat $\phi$CDM model is most favored while the non-flat $\phi$CDM model is least favored. However,
given the issue raised above about the 2019 QSO data, it is inappropriate to give much weight to these findings.

The cosmological parameters obtained by using the 2019 QSO data have relatively high uncertainty for all models so they are mostly consistent with the results obtained by using the  BAO + $H(z)$ data set, as can be seen from Figs.\ 4--15.

\subsection{QSO + $H(z)$ + BAO constraints}
\label{QSOBAOH2}
Results for the $H(z)$ + BAO observations are given in Tables 3--5 and one-dimensional distributions and two-dimensional contours are shown in red in Figs. \ 4--15. Figures 4--15 show that constraints from the QSO data alone and those from the BAO + $H(z)$ data are mostly consistent with each other. So it is not unresonable to do joint analyses of the QSO + $H(z)$ +BAO data. Results from this joint analysis are listed in Tables 1, 2, and 6. The QSO + $H(z)$ + BAO one-dimensional likelihoods and two-dimensional confidence contours for all free parameters are shown in blue in Figs.\ 4--15. The updated QSO data don't significantly tighten the BAO + $H(z)$ data contours except in the cases of the non-flat XCDM parametrization and the non-flat $\phi$CDM model (Figs.\ 10, 11, 14, and 15). Another noticeable result is that adding the QSO data to the BAO + $H(z)$ data results in the shifting of one-dimensional likelihood distribution of the matter density parameter towards higher values in most cosmological models studied here.

From joint analyses of the QSO + $H(z)$ + BAO data, from Table 6, the matter density parameter lies in the range $\om$ = $0.30 \pm 0.02$ to $0.31 \pm 0.01$ ($\om$ = $0.30 \pm 0.01$ to $0.33 \pm 0.02$)  for flat (non-flat) models and the $H_0 = 68 \pm 2.8$ ${\rm km}\hspace{1mm}{\rm s}^{-1}{\rm Mpc}^{-1}$ prior and lies in the range $\om$ = $0.30^{+0.02}_{-0.01}$ to $0.31 \pm 0.01$ ($\om$ = $0.31 \pm 0.01$ to $0.33 \pm 0.02$)  for flat (non-flat) models and the $H_0 = 73.24 \pm 1.74$ ${\rm km}\hspace{1mm}{\rm s}^{-1}{\rm Mpc}^{-1}$ prior. In a few cases these results slightly differ from the BAO + $H(z)$ data results in Table 5, being shifted to slightly larger values. These results are consistent with those results determined from other cosmological data.

The Hubble constant lies in the range $H_0$ = $66.76^{+1.36}_{-1.49}$ to $68.04^{+0.84}_{-0.84}$ ($H_0$ = $67.48^{+1.81}_{-1.77}$ to $68.70^{+1.78}_{-1.79}$) ${\rm km}\hspace{1mm}{\rm s}^{-1}{\rm Mpc}^{-1}$ for flat (non-flat) models and the $H_0 = 68 \pm 2.8$ ${\rm km}\hspace{1mm}{\rm s}^{-1}{\rm Mpc}^{-1}$ prior and lies in the range $H_0$ = $69.09^{+1.01}_{-1.02}$ to $71.38^{+1.51}_{-1.50}$ ($H_0$ = $71.17^{+1.45}_{-1.43}$ to $71.79^{+1.40}_{-1.39}$) ${\rm km}\hspace{1mm}{\rm s}^{-1}{\rm Mpc}^{-1}$ for flat (non-flat) models and the $H_0 = 73.24 \pm 1.74$ ${\rm km}\hspace{1mm}{\rm s}^{-1}{\rm Mpc}^{-1}$ prior. Not unexpectedly, for the $H_0 = 73.24 \pm 1.74$ ${\rm km}\hspace{1mm}{\rm s}^{-1}{\rm Mpc}^{-1}$ prior case the measured value of $H_0$ is reduced below the prior value because the BAO and $H(z)$ observations prefer a lower $H_0$. In most cases the $H_0$ error bars have increased in comparison to those derived using the 2015 QSO + BAO + $H(z)$ data in \cite{Khadka2019}.

In all models, except for non-flat $\Lambda$CDM with the $H_0 = 68 \pm 2.8$ ${\rm km}\hspace{1mm}{\rm s}^{-1}{\rm Mpc}^{-1}$ prior, closed spatial hypersurfaces are favored at about 2$\sigma$. For the non-flat $\Lambda$CDM model the curvature energy density parameter is $\ok$ = $-0.01^{+0.06}_{-0.07}$ and $-0.09 \pm 0.05$ for the $H_0 = 68 \pm 2.8$ ${\rm km}\hspace{1mm}{\rm s}^{-1}{\rm Mpc}^{-1}$ and $73.24 \pm 1.74$ ${\rm km}\hspace{1mm}{\rm s}^{-1}{\rm Mpc}^{-1}$ priors respectively. Values of curvature energy density parameter obtained for non-flat dynamical dark energy cosmological models are significantly higher than those obtained in the non-flat $\Lambda$CDM model. The curvature energy density parameter is $\ok$ = $-0.34 \pm 0.18$ and $-0.32 \pm 0.16$ for the non-flat XCDM and non-flat $\phi$CDM models for the $H_0 = 68 \pm 2.8$ ${\rm km}\hspace{1mm}{\rm s}^{-1}{\rm Mpc}^{-1}$ prior and $\ok$ = $-0.31^{+0.17}_{-0.18}$ and $-0.35 \pm 0.15$ for the non-flat XCDM and non-flat $\phi$CDM models for the $H_0 = 73.24 \pm 1.74$ ${\rm km}\hspace{1mm}{\rm s}^{-1}{\rm Mpc}^{-1}$ prior. This preference for closed spatial geometries is largely driven by the BAO + $H(z)$ data \citep{Park2018d, Ryan2019}.

From Table 6, for the flat (non-flat) $\Lambda$CDM model the dark energy density parameter is $\ol$ = $0.70 \pm 0.01$ ($0.71^{+0.05}_{-0.06}$) for the $H_0 = 68 \pm 2.8$ ${\rm km}\hspace{1mm}{\rm s}^{-1}{\rm Mpc}^{-1}$ prior and $\ol$ = $0.69 \pm 0.01$ ($0.78 \pm 0.04$) for the $H_0 = 73.24 \pm 1.74$ ${\rm km}\hspace{1mm}{\rm s}^{-1}{\rm Mpc}^{-1}$ prior.

The equation of state parameter for the flat (non-flat) XCDM parametrization is $\omega_{X}$ = $-0.96 \pm 0.09$ ($-0.69^{+0.07}_{-0.11}$) for the $H_0 = 68 \pm 2.8$ ${\rm km}\hspace{1mm}{\rm s}^{-1}{\rm Mpc}^{-1}$ prior and $-1.14 \pm 0.08$ ($-0.77^{+0.09}_{-0.15}$) for the $73.24 \pm 1.74$ ${\rm km}\hspace{1mm}{\rm s}^{-1}{\rm Mpc}^{-1}$ prior. So this set of data suggests decreasing XCDM dark energy density with time, except for the flat XCDM parametrization with $73.24 \pm 1.74$ ${\rm km}\hspace{1mm}{\rm s}^{-1}{\rm Mpc}^{-1}$ prior, where it favors at almost 2$\sigma$, a XCDM dark energy density that increases with time. The value of the $\alpha$ parameter in the flat (non-flat) $\phi$CDM model is $\alpha$ = $0.20^{+0.21}_{-0.14}$ ($1.21^{+0.47}_{-0.53}$) for the $H_0 = 68 \pm 2.8$ ${\rm km}\hspace{1mm}{\rm s}^{-1}{\rm Mpc}^{-1}$ prior and $0.06^{+0.09}_{-0.05}$ ($0.98^{+0.44}_{-0.50}$) for the $73.24 \pm 1.74$ ${\rm km}\hspace{1mm}{\rm s}^{-1}{\rm Mpc}^{-1}$ prior. All eight XCDM and $\phi$CDM cases favor dynamical dark energy over a $\Lambda$ at between 0.4$\sigma$ and 4.4$\sigma$. Other data also favor mild dark energy dynamics \citep{Ooba2018d, Park2018b, Park2018c}.

Unlike the case for the 2019 QSO only data, for the QSO + BAO $H(z)$ data the $\chi^2_{\rm min}$, AIC, and BIC values are relatively similar for all models.

\section{Conclusion}
\label{con}
Following \cite{Risaliti2019} we have used the correlation between X-ray and UV monochromatic luminosities in selected $z \sim 0 - 5$ quasars to constrain cosmological model parameters in six different models. These selected quasars can be used as standard candles for cosmological model testing at redshifts $z \sim 2.5 - 5$ that are not yet widely accessible through other cosmological probes. Our analyses of these data in six different cosmological models shows that parameters of the $L_{X}-L_{UV}$ relation, i.e., the intercept $\beta$, the slope $\gamma$, and the intrinsic dispersion $\delta$, are only weakly dependent on the cosmological model assumed in the analysis. This reinforces the finding of \cite{Risaliti2015} that carefully-selected quasar flux measurements can be used as standard candles.

The 2019 QSO data constraints are mostly consistent with joint analysis of BAO distance and Hubble parameter measurements, as also found in \cite{Khadka2019} for the 2015 QSO data. We find that combined analysis of 2019 QSO and BAO + $H(z)$ measurements slightly tightens the $H(z)$ + BAO data constraints in the non-flat XCDM paramerization and the non-flat $\phi$CDM model but not in the other four models. Overall, adding the 2019 QSO measurements to the BAO + $H(z)$ observations has a less significant tightening effect than what was found for the 2015 QSO data \citep{Khadka2019}.

The value of the matter density parameter obtained by using the 2019 QSO data is typically greater than 0.5, Tables 1 and 2, which is significantly larger than values obtained using other cosmological probes, such as BAO, $H(z)$, Type Ia supernovae, and CMB anisotropy observations. Due to the larger $\om$ from the QSO data, in joint analyses of the QSO + BAO + $H(z)$ measurements the matter density parameter shifts to slightly larger values than the $H(z)$ + BAO data $\om$ values in a number of the models. The larger 2019 QSO data $\om$ values are likely the cause of the tension between the 2019 QSO data and the $\om = 0.3$ flat $\Lambda$CDM model that is discussed in \cite{Risaliti2019} and \cite{Lusso2019}. It is probably more likely that this tension has to do with the $z \sim 2 - 5$ 2019 QSO data than with the invalidity of the $\om = 0.3$ flat $\Lambda$CDM model. This is because almost all cosmological data, at $z \sim 0 - 2.5$ and at $z \sim 1100$, are consistent with $\om \cong 0.3$. It is of great interest to understand why the 2019 QSO observations favour a larger value of $\om$.

The joint QSO + BAO + $H(z)$ measurements constraints are consistent with the current standard spatially-flat $\Lambda$CDM model, but weakly favour dynamical dark energy over a cosmological constant and closed over flat spatial hypersurfaces. Since they probe a little-studied, higher redshift region of the universe, future, improved QSO data will likely provide very useful, more restrictive, constraints on cosmological parameters, and should help to measure the dynamics of dark energy and the geometry of space.

\section*{Acknowledgements}
We thank Elisabeta Lusso for her generous help, and Lado Samushia, Javier De Cruz, Shulei Cao, and Joe Ryan for useful discussions. We are grateful to the Beocat Research Cluster at Kansas State University team. This research was supported in part by DOE grant DE-SC0019038.
\section*{Data availability}
The data underlying this article were provided to us by the authors of \cite{Risaliti2019}. These data will be shared on request to the corresponding author with the permission of the authors of \cite{Risaliti2019}.

\begin{figure*}
\begin{multicols}{2}
    \includegraphics[width=\linewidth]{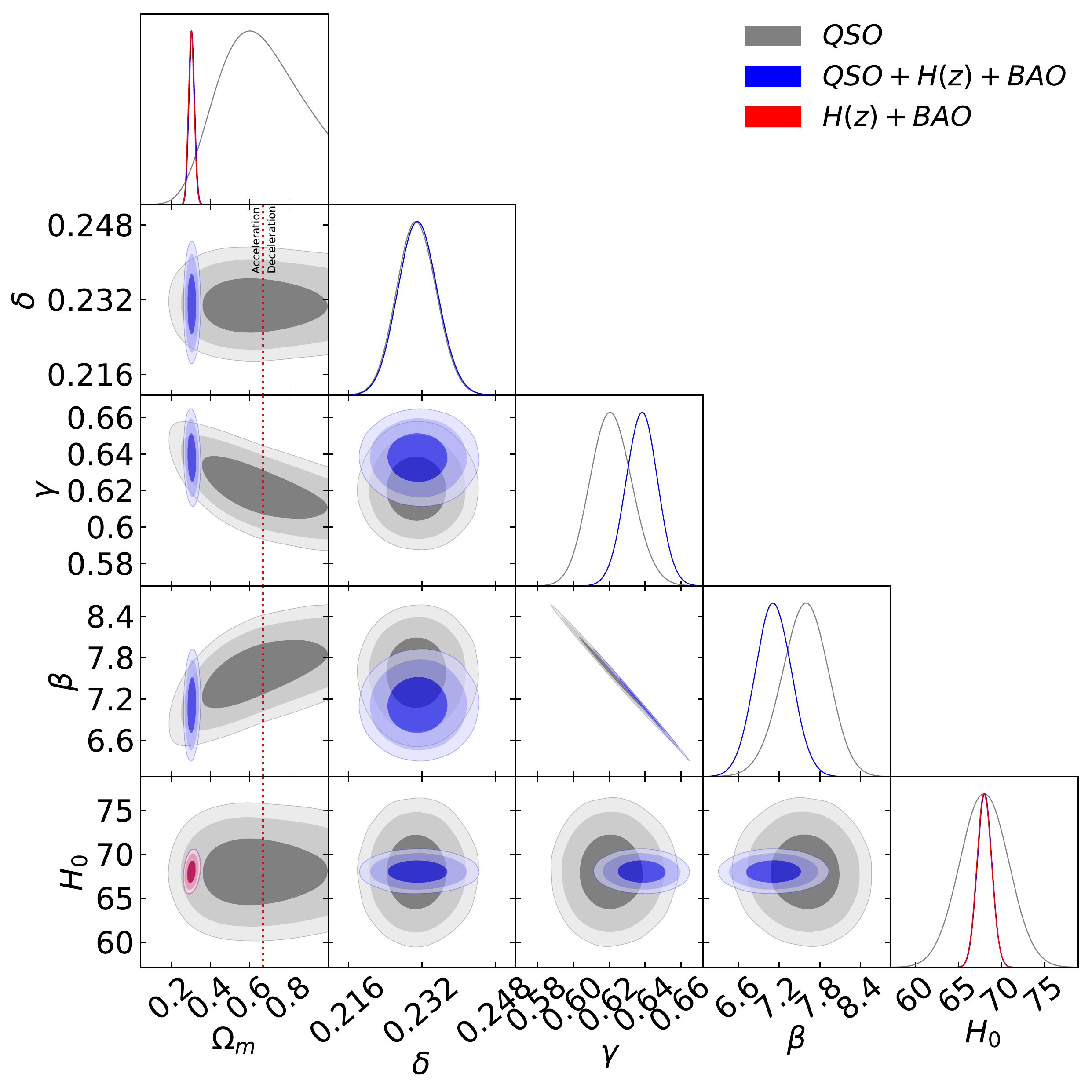}\par
    \includegraphics[width=\linewidth]{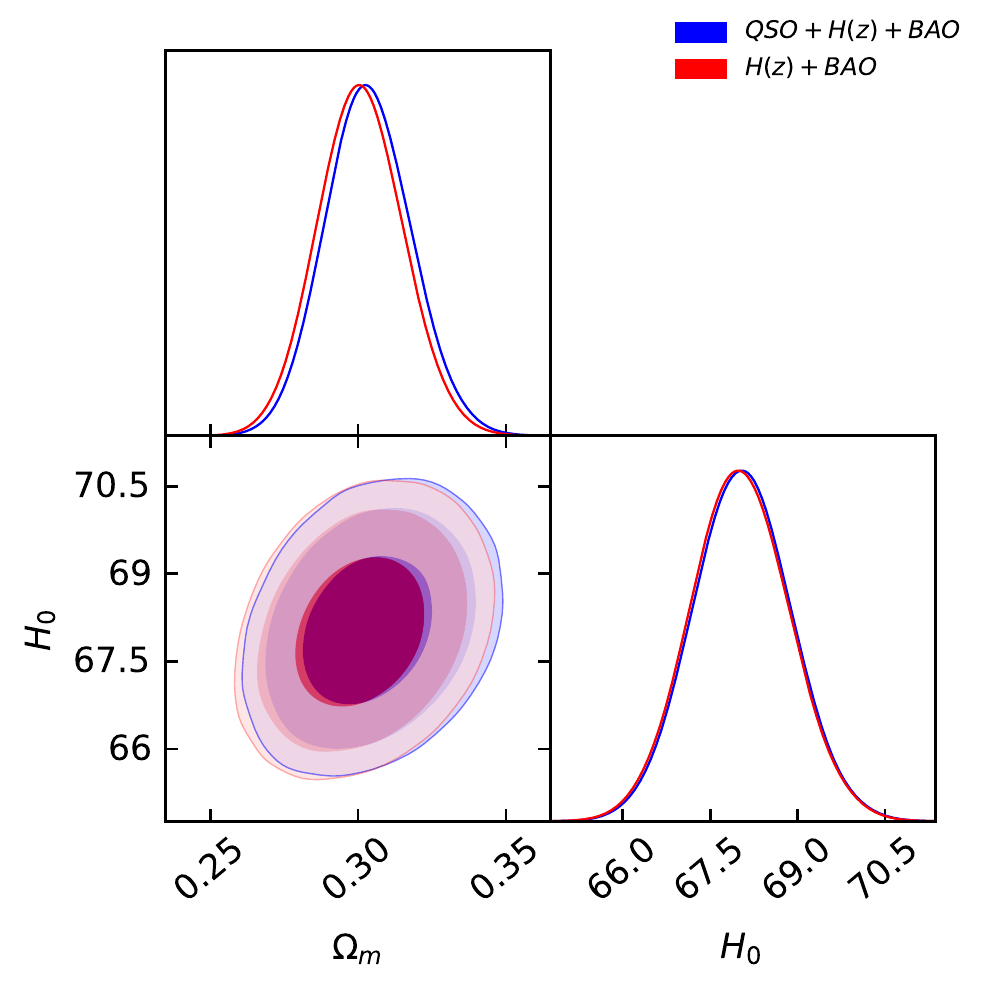}\par
\end{multicols}
\caption{Flat \lcdm\ model constraints from QSO (grey), $H(z)$ + BAO (red),  and QSO + $H(z)$ + BAO (blue) data. Left panel shows 1, 2, and 3$\sigma$ confidence contours and one-dimensional likelihoods for all free parameters. The red dotted straight lines are zero acceleration lines, with currently accelerated cosmological expansion occurring to the left of the lines. Right panel shows magnified plots for only cosmological parameters $\om$ and $H_0$, without the QSO-only constraints. These plots are for the $H_0 = 68 \pm 2.8$ ${\rm km}\hspace{1mm}{\rm s}^{-1}{\rm Mpc}^{-1}$ prior.}
\label{fig:flat LCDM68 model with BAO, H(z) and QSO data}
\end{figure*}
\begin{figure*}
\begin{multicols}{2}
    \includegraphics[width=\linewidth]{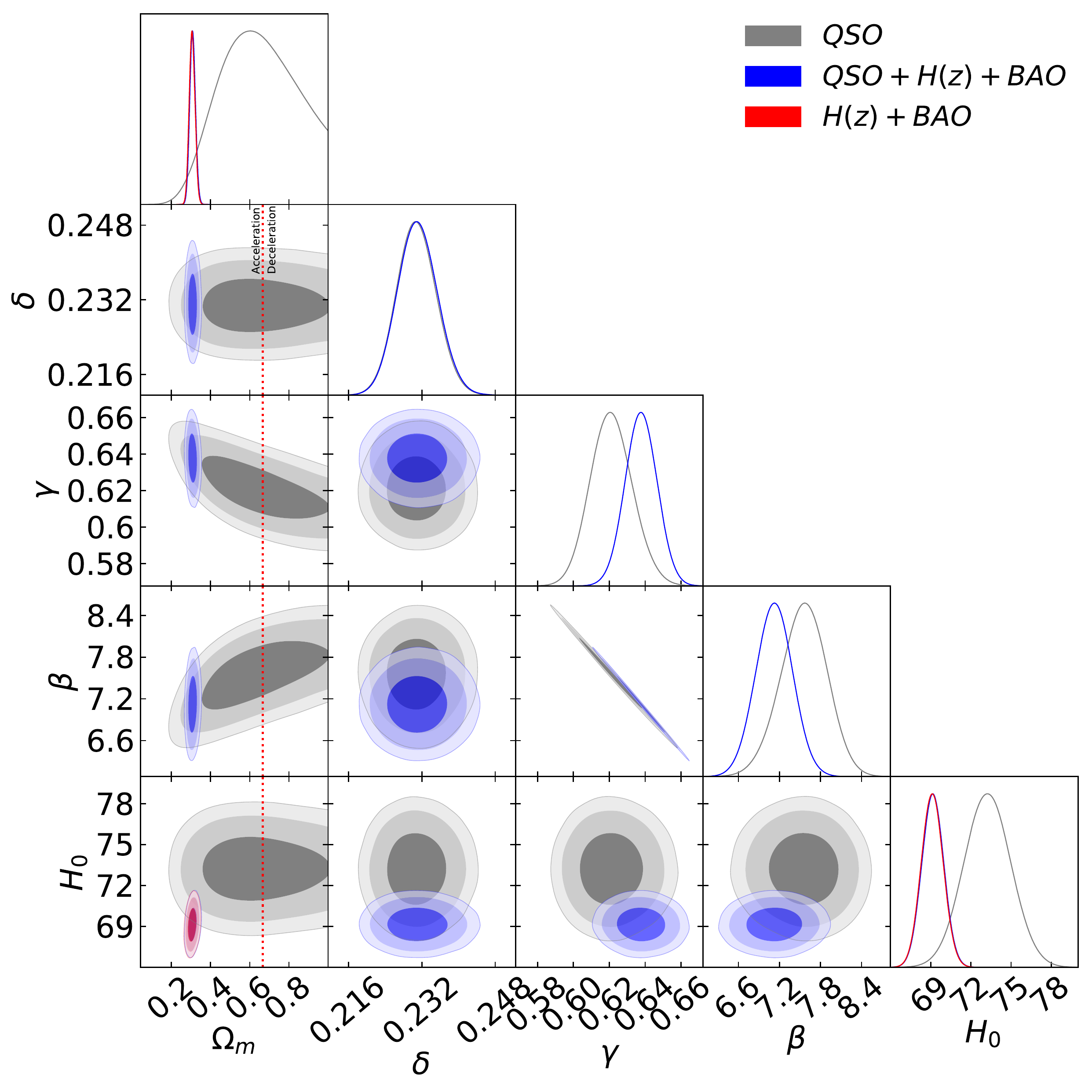}\par
    \includegraphics[width=\linewidth]{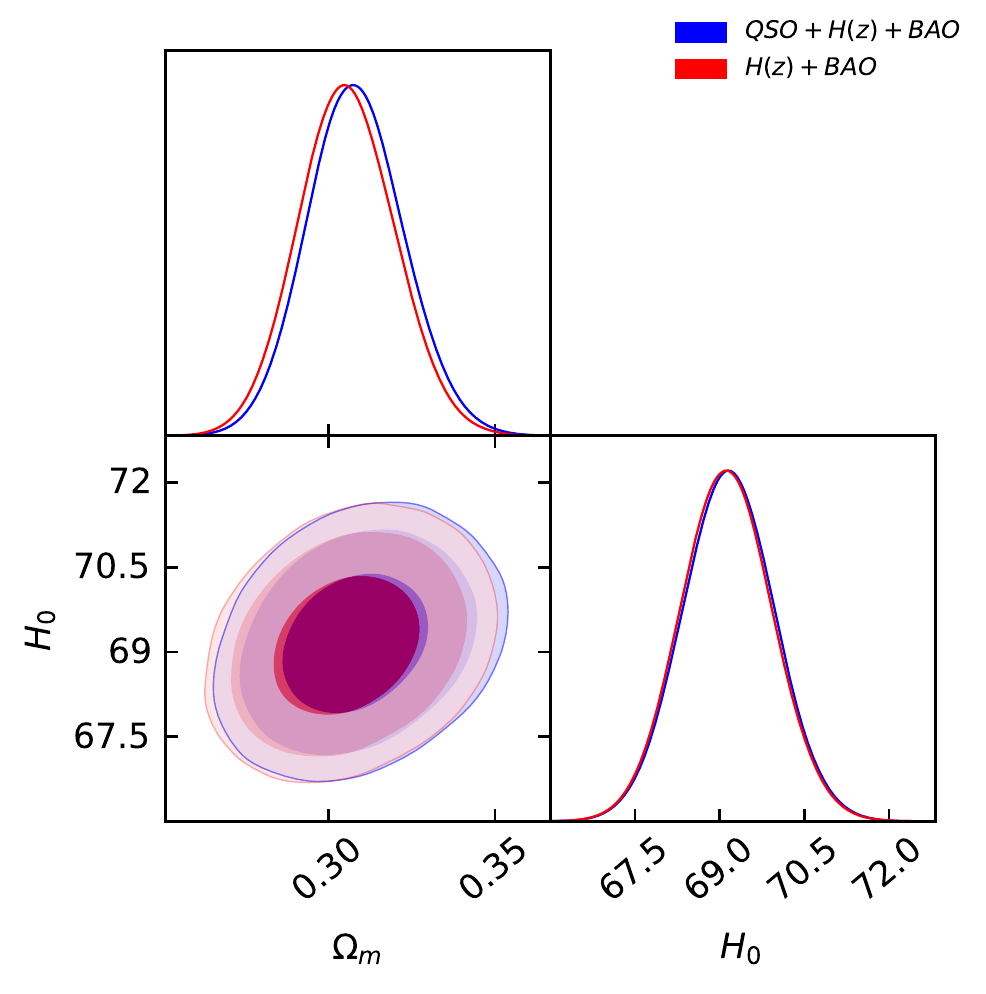}\par
\end{multicols}
\caption{Flat \lcdm\ model constraints from QSO (grey), $H(z)$ + BAO (red),  and QSO + $H(z)$ + BAO (blue) data. Left panel shows 1, 2, and 3$\sigma$ confidence contours and one-dimensional likelihoods for all free parameters. The red dotted straight lines are zero acceleration lines, with currently accelerated cosmological expansion occurring to the left of the lines. Right panel shows magnified plots for only cosmological parameters $\om$ and $H_0$, without the QSO-only constraints. These plots are for the $H_0 = 73.24 \pm 1.74$ ${\rm km}\hspace{1mm}{\rm s}^{-1}{\rm Mpc}^{-1}$ prior.}
\label{fig:flat LCDM73 model with BAO, H(z) and QSO data}
\end{figure*}
\begin{figure*}
\begin{multicols}{2}
    \includegraphics[width=\linewidth]{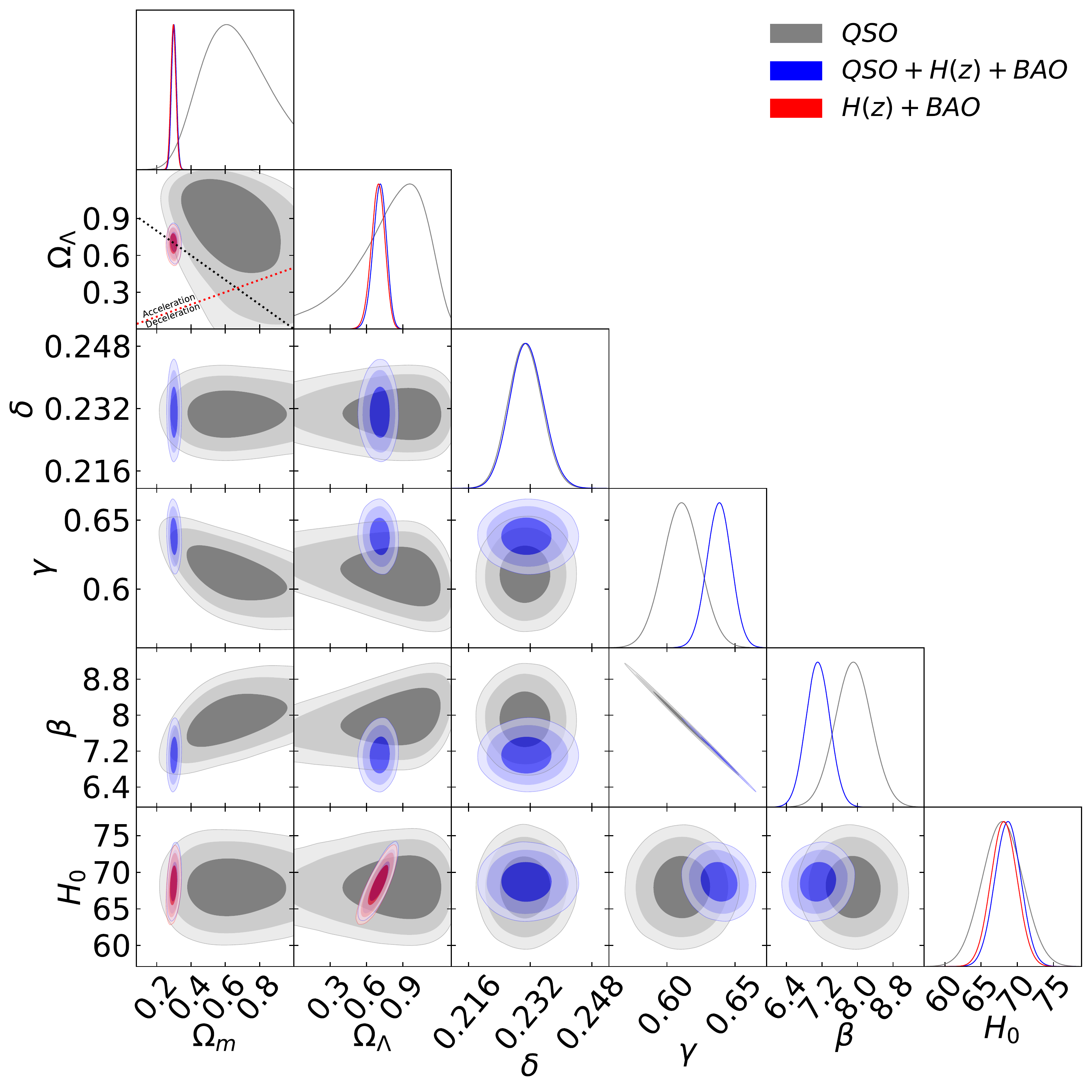}\par
    \includegraphics[width=\linewidth]{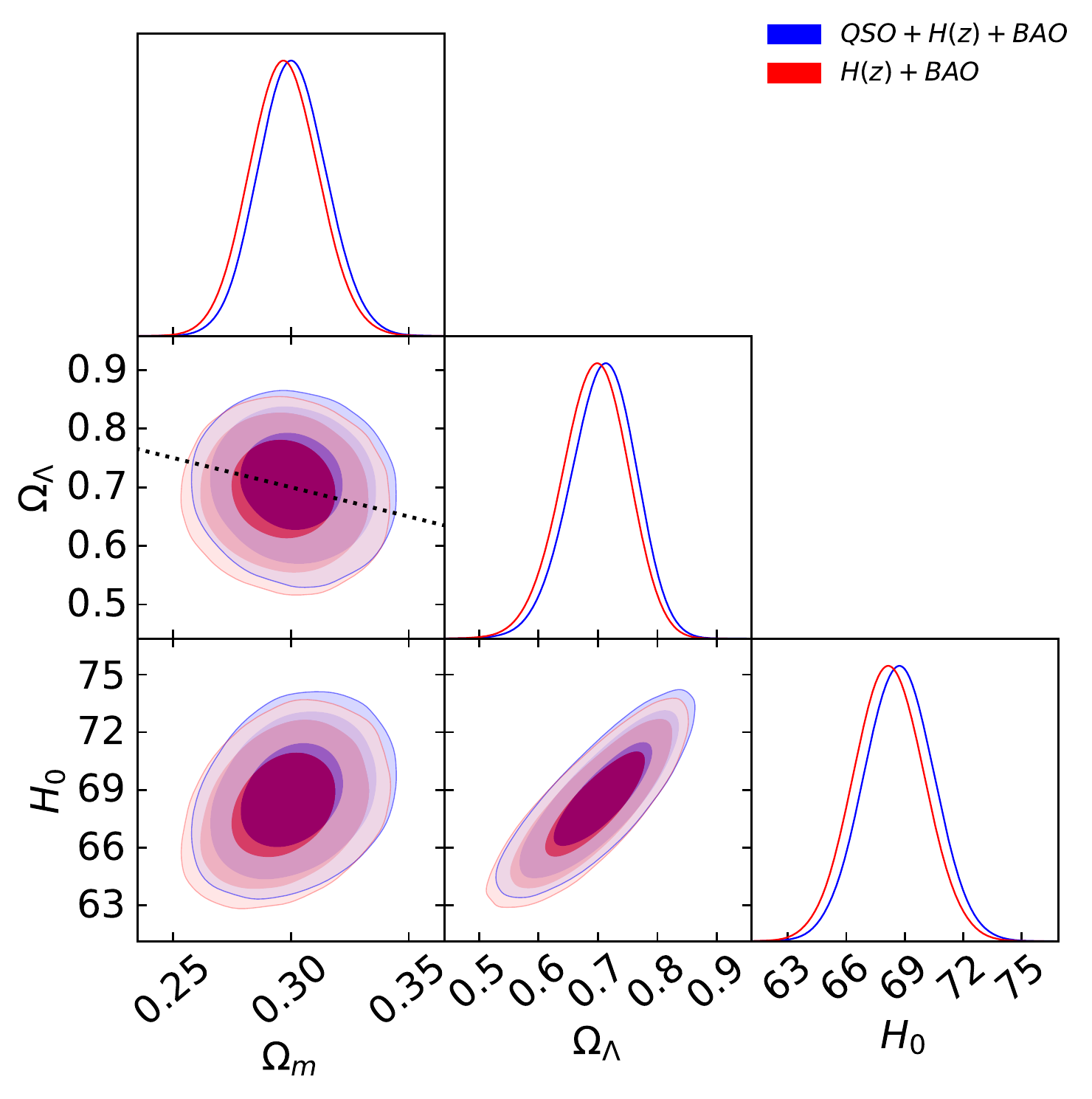}\par
\end{multicols}
\caption{Non-flat \lcdm\ model constraints from QSO (grey), $H(z)$ + BAO (red),  and QSO + $H(z)$ + BAO (blue) data. Left panel shows 1, 2, and 3$\sigma$ confidence contours and one-dimensional likelihoods for all free parameters. The red dotted straight line in the $\ol - \om $ panel is the zero acceleration line with currently accelerated cosmological expansion occurring to the upper left of the line. Right panel shows magnified plots for cosmological parameters $\om$, $\ol$, and $H_0$, without the QSO-only constraints. These plots are for the $H_0 = 68 \pm 2.8$ ${\rm km}\hspace{1mm}{\rm s}^{-1}{\rm Mpc}^{-1}$ prior. The black dotted straight line in the $\ol - \om $ panels correspond to the flat $\Lambda$CDM model, with closed spatial geometry being to the upper right.}
\label{fig:non-flat LCDM68 model with BAO, H(z) and QSO data}
\end{figure*}
\begin{figure*}
\begin{multicols}{2}
    \includegraphics[width=\linewidth]{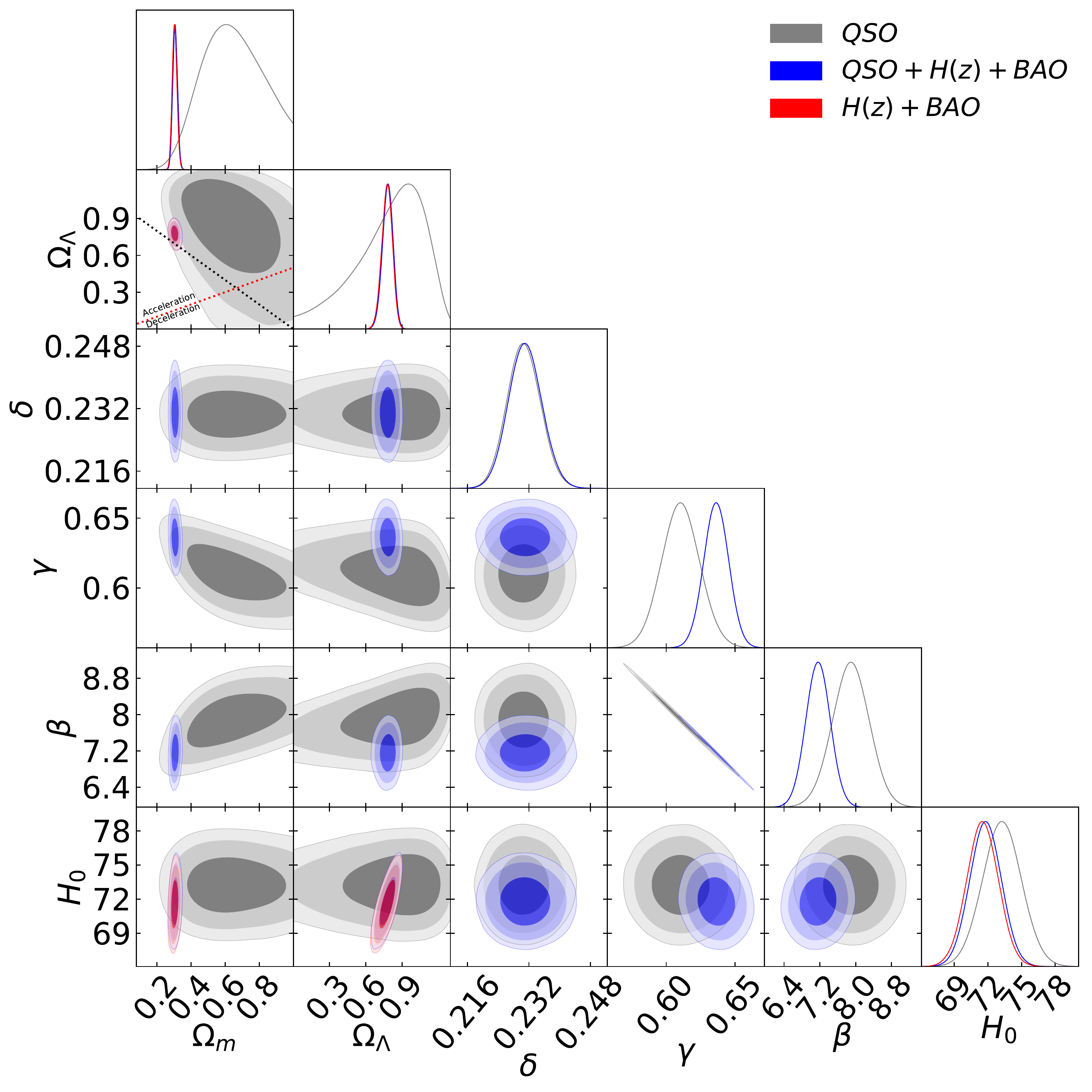}\par
    \includegraphics[width=\linewidth]{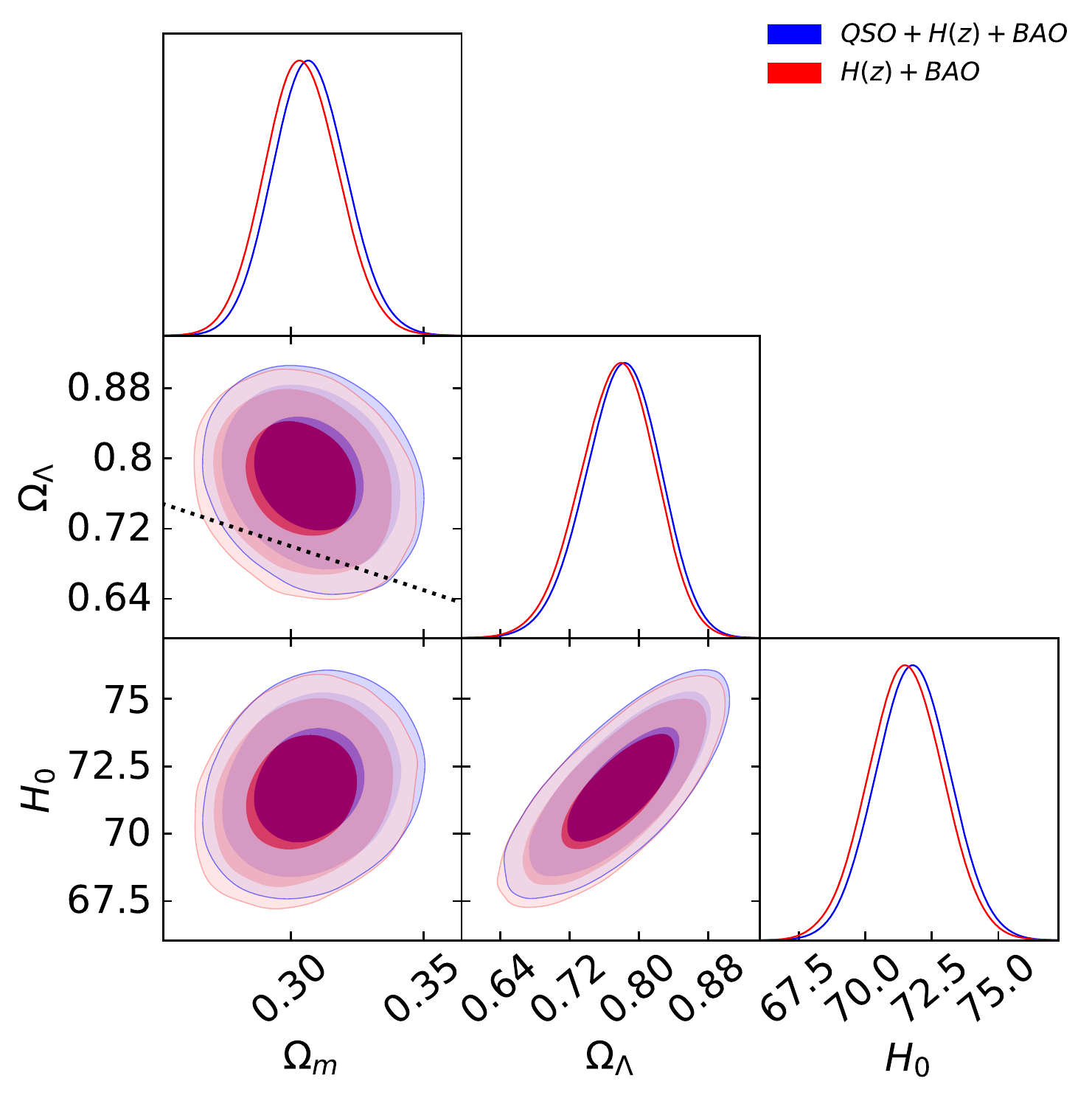}\par
\end{multicols}
\caption{Non-flat \lcdm\ model constraints from QSO (grey), $H(z)$ + BAO (red),  and QSO + $H(z)$ + BAO (blue) data. The red dotted straight line in the $\ol - \om $ panel is the zero acceleration line with currently accelerated cosmological expansion occurring to the upper left of the line. Left panel shows 1, 2, and 3$\sigma$ confidence contours and one-dimensional likelihoods for all free parameters. Right panel shows magnified plots for only cosmological parameters $\om$, $\ol$, and $H_0$, without the QSO-only constraints. These plots are for the $H_0 = 73.24 \pm 1.74$ ${\rm km}\hspace{1mm}{\rm s}^{-1}{\rm Mpc}^{-1}$ prior. The black dotted straight line in the $\ol - \om $ panels correspond to the flat $\Lambda$CDM model, with closed spatial geometry being to the upper right.}
\label{fig:non-flat LCDM73 model with BAO, H(z) and QSO data}
\end{figure*}
\begin{figure*}
\begin{multicols}{2}
    \includegraphics[width=\linewidth]{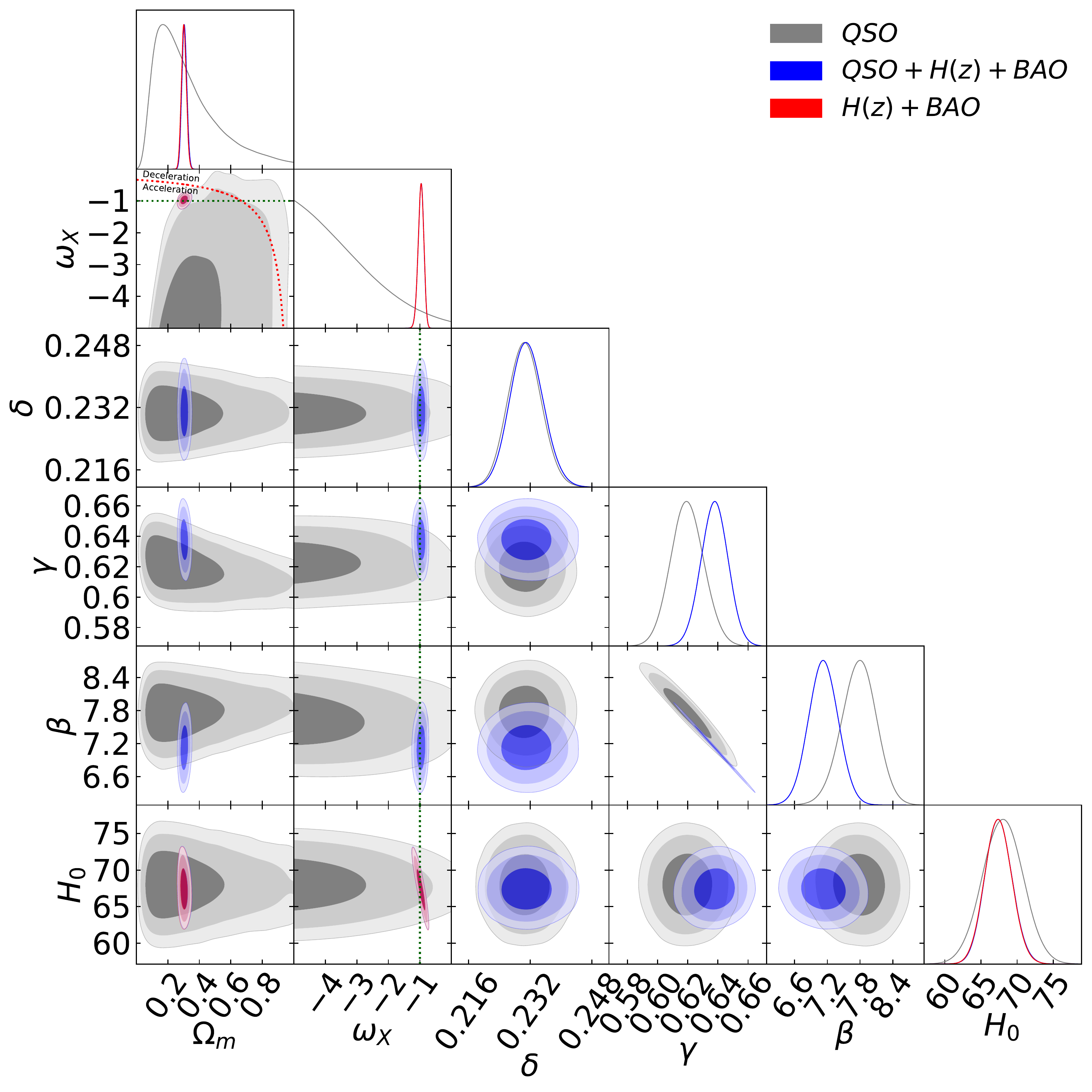}\par
    \includegraphics[width=\linewidth]{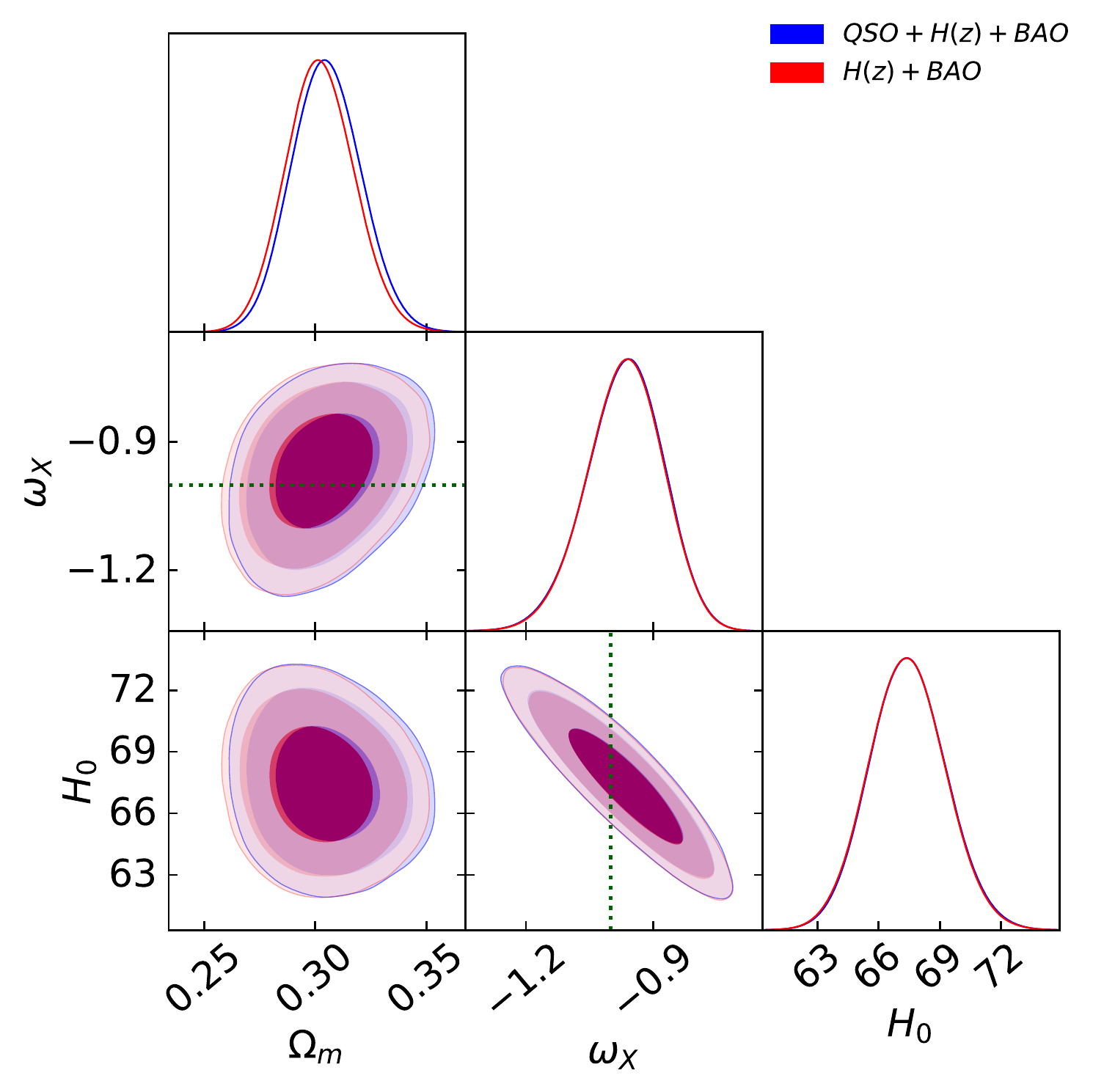}\par
\end{multicols}
\caption{Flat XCDM parametrization constraints from QSO (grey), $H(z)$ + BAO (red),  and QSO + $H(z)$ + BAO (blue) data. Left panel shows 1, 2, and 3$\sigma$ confidence contours and one-dimensional likelihoods for all free parameters. The red dotted curved line in the $\omega_X - \om$ panel is the zero acceleration line with currently accelerated cosmological expansion occurring below the line. Right panel shows magnified plots for only cosmological parameters $\om$, $\omega_X$, and $H_0$, without the QSO-only constraints. These plots are for the $H_0 = 68 \pm 2.8$ ${\rm km}\hspace{1mm}{\rm s}^{-1}{\rm Mpc}^{-1}$ prior. The green dotted straight lines represent $\omega_x$ = $-1$.}
\label{fig:Flat XCDM68 model with BAO, H(z) and QSO data}
\end{figure*}
\begin{figure*}
\begin{multicols}{2}
    \includegraphics[width=\linewidth]{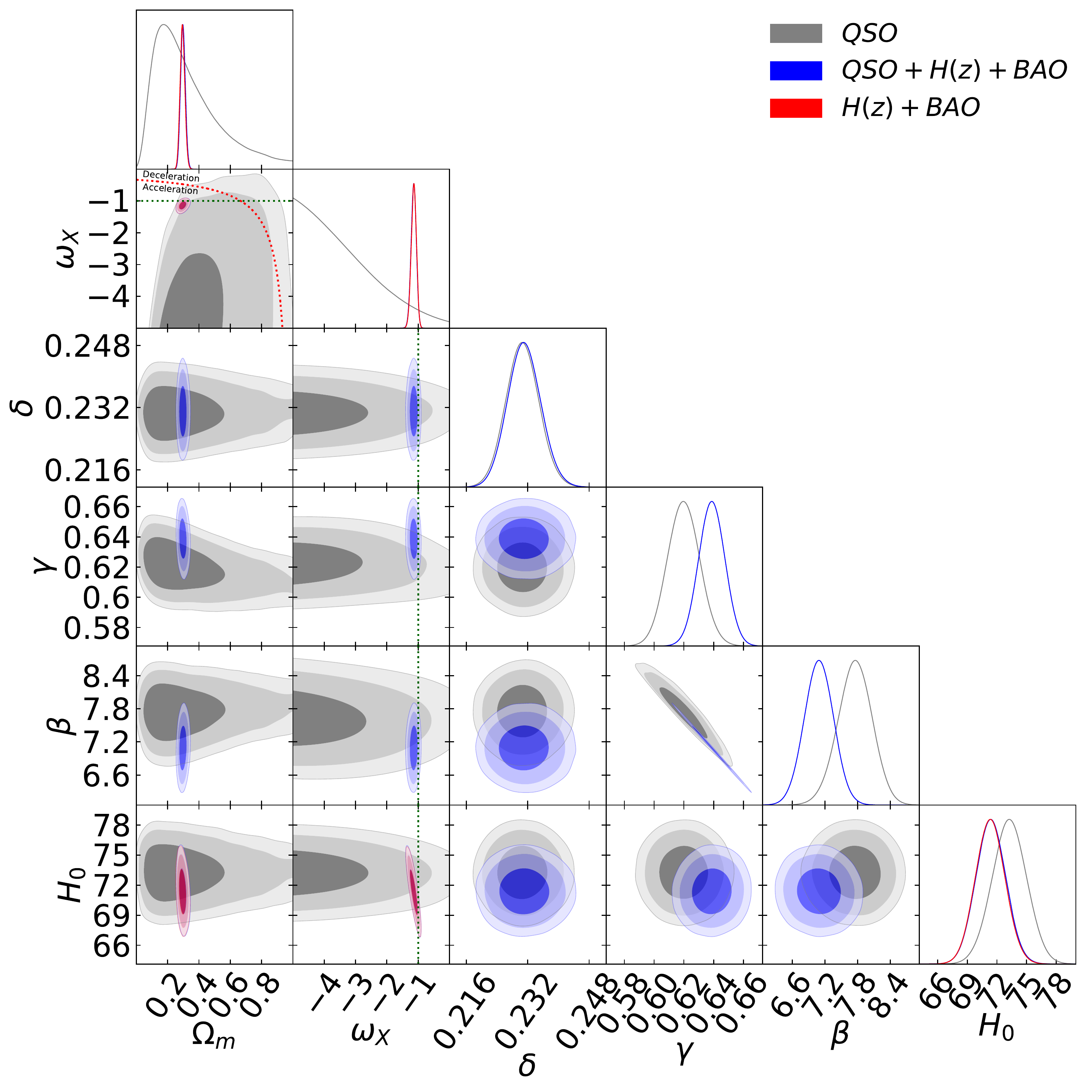}\par
    \includegraphics[width=\linewidth]{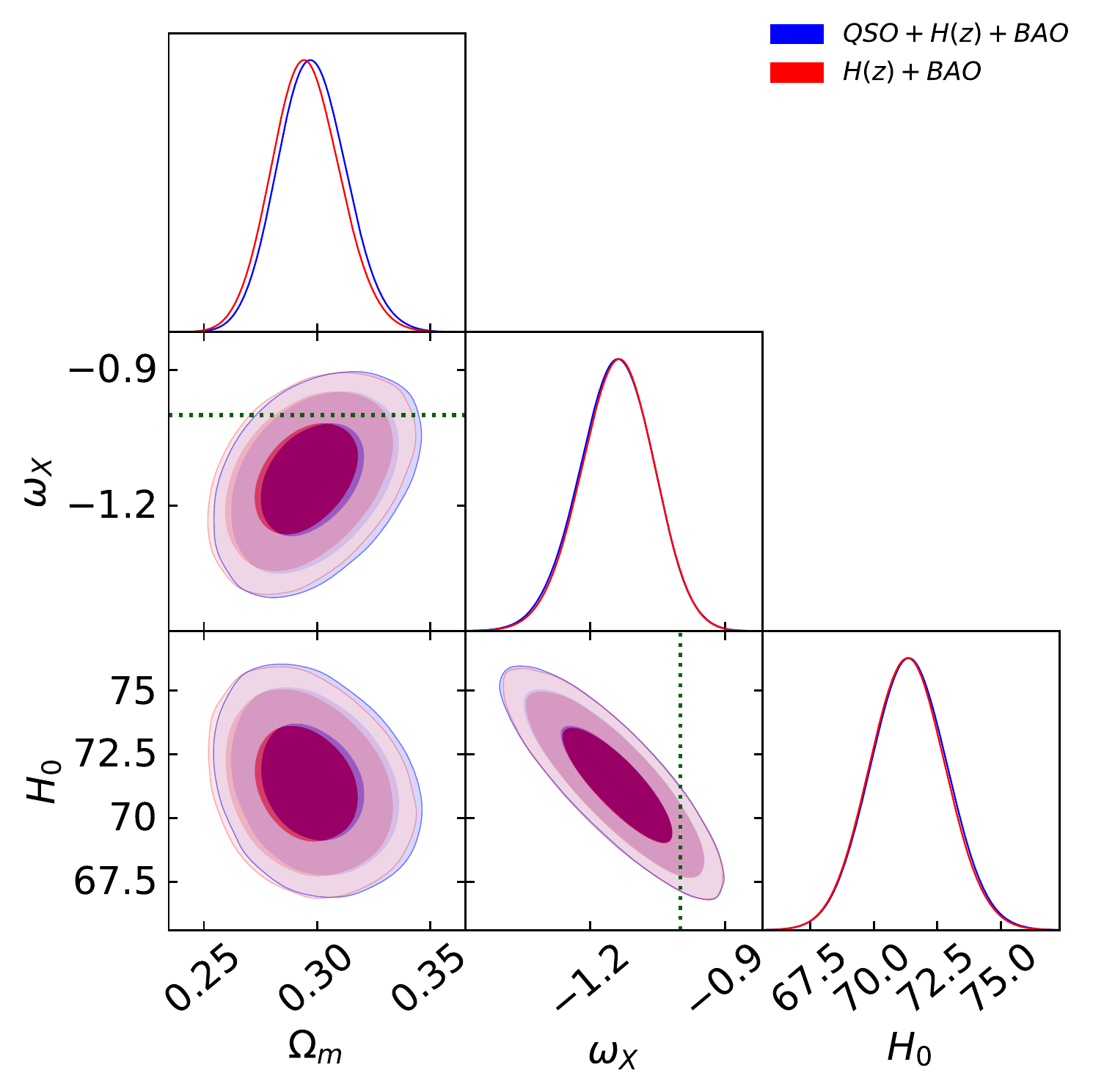}\par
\end{multicols}
\caption{Flat XCDM parametrization constraints from QSO (grey), $H(z)$ + BAO (red),  and QSO + $H(z)$ + BAO (blue) data. Left panel shows 1, 2, and 3$\sigma$ confidence contours and one-dimensional likelihoods for all free parameters. The red dotted curved line in the $\omega_X - \om$ panel is the zero acceleration line with currently accelerated cosmological expansion occurring below the line. Right panel shows magnified plots for only cosmological parameters $\om$, $\omega_X$, and $H_0$, without the QSO-only constraints. These plots are for the $H_0 = 73.24 \pm 1.74$ ${\rm km}\hspace{1mm}{\rm s}^{-1}{\rm Mpc}^{-1}$ prior. The green dotted straight lines represent $\omega_x$ = $-1$.}
\label{fig:Flat XCDM73 model with BAO, H(z) and QSO data}
\end{figure*}
\begin{figure*}
\begin{multicols}{2}
    \includegraphics[width=\linewidth]{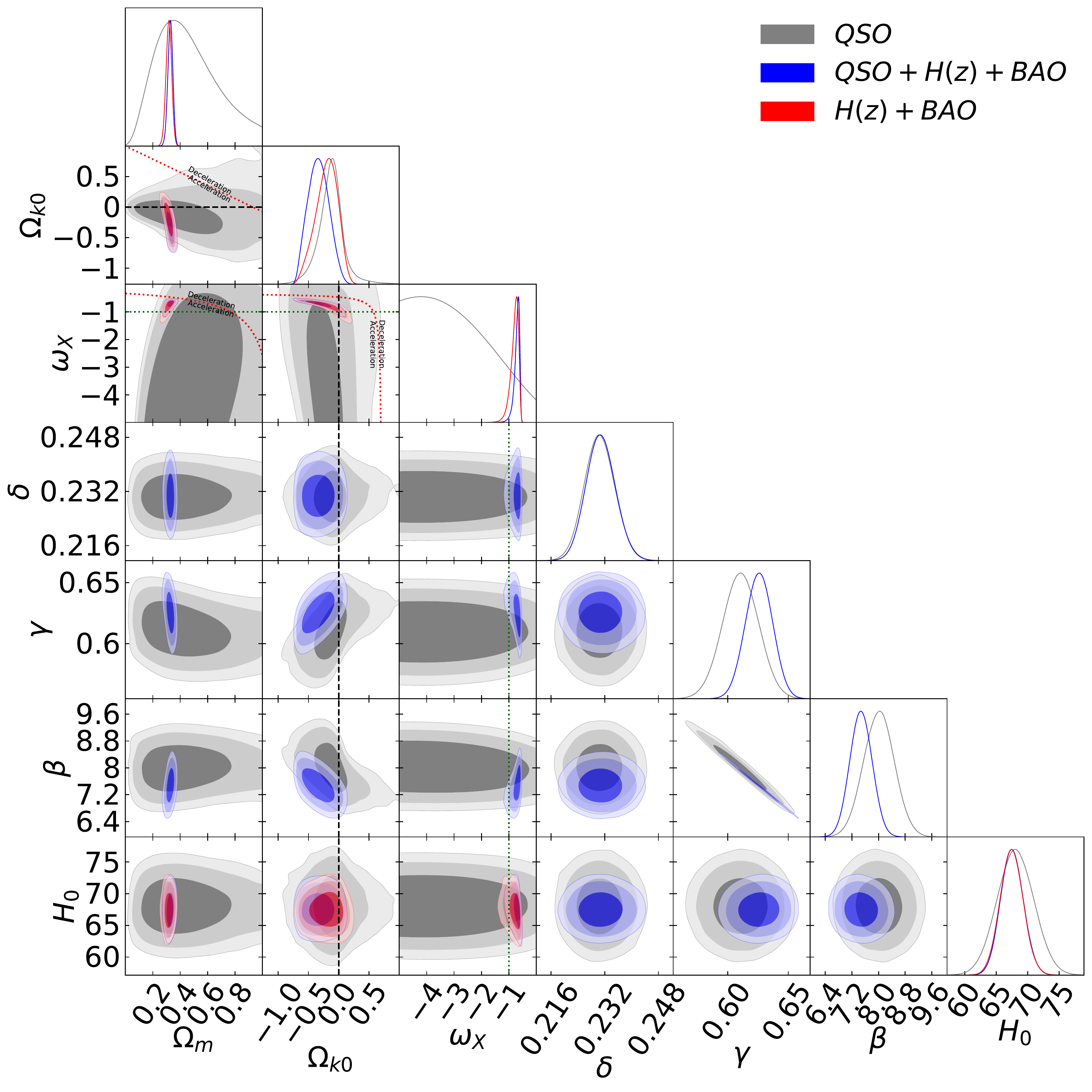}\par
    \includegraphics[width=\linewidth]{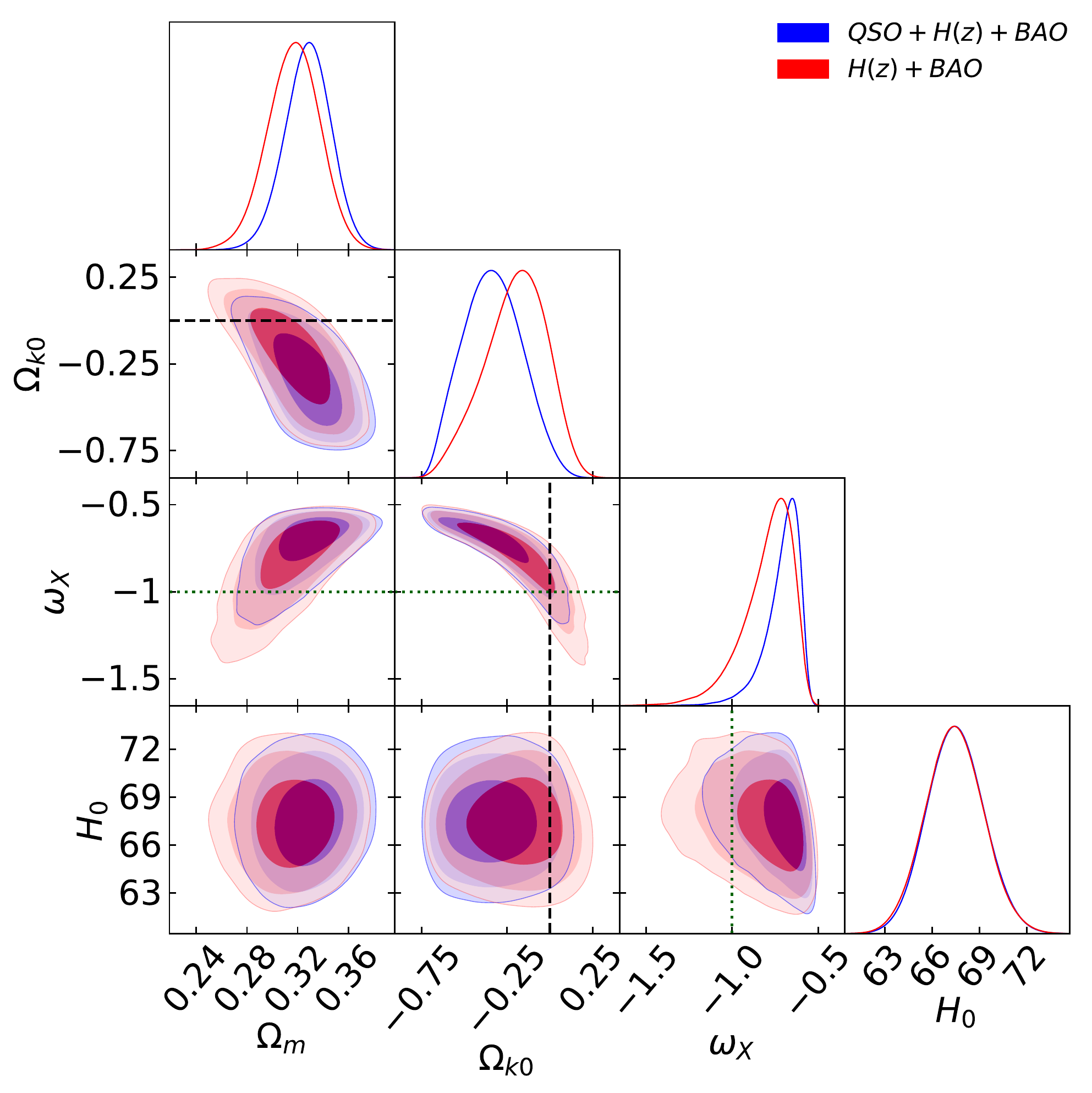}\par
\end{multicols}
\caption{Non-flat XCDM parametrization constraints from QSO (grey), $H(z)$ + BAO (red),  and QSO + $H(z)$ + BAO (blue) data. Left panel shows 1, 2, and 3$\sigma$ confidence contours and one-dimensional likelihoods for all free parameters. The red dotted curved lines in the $\omega_{K0} - \om$, $\omega_X - \om$, and $\omega_X - \Omega_{k0}$
panels are the zero acceleration lines with currently accelerated cosmological expansion occurring below the lines. Each of the three lines are computed with the third parameter set to the QSO data only best-fit value of Table 3. Right panel shows magnified plots for only cosmological parameters $\om$, $\ok$, $\omega_X$, and $H_0$, without the QSO-only constraints. These plots are for the $H_0 = 68 \pm 2.8$ ${\rm km}\hspace{1mm}{\rm s}^{-1}{\rm Mpc}^{-1}$ prior. The black dashed straight lines and the green dotted straight lines are $\ok$ = 0 and $\omega_x$ = $-1$ lines.}
\label{fig:non-flat XCDM68 model with BAO, H(z) and QSO data}
\end{figure*}
\begin{figure*}
\begin{multicols}{1}
    \includegraphics[width=\linewidth]{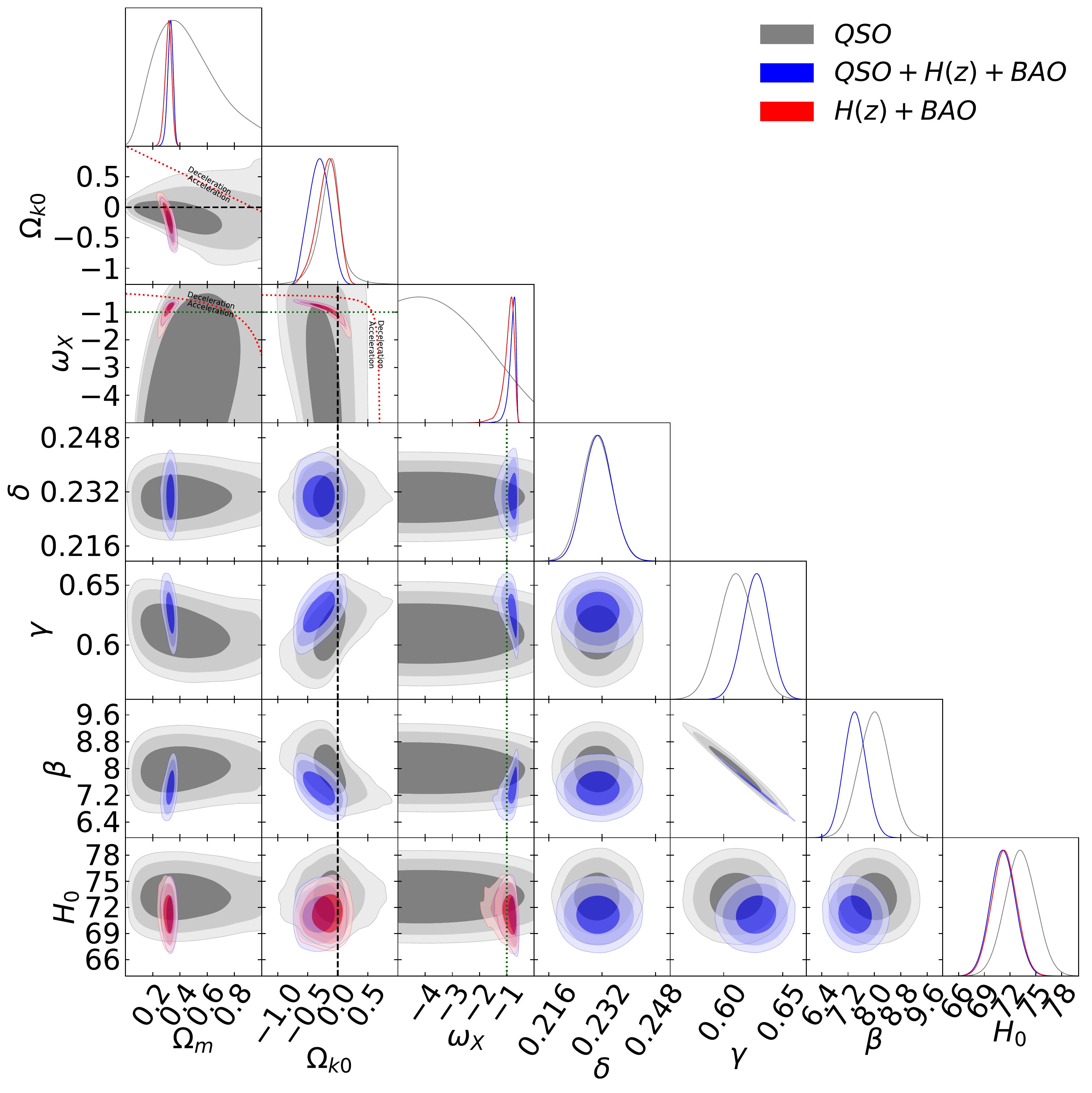}\par
    \includegraphics[width=\linewidth]{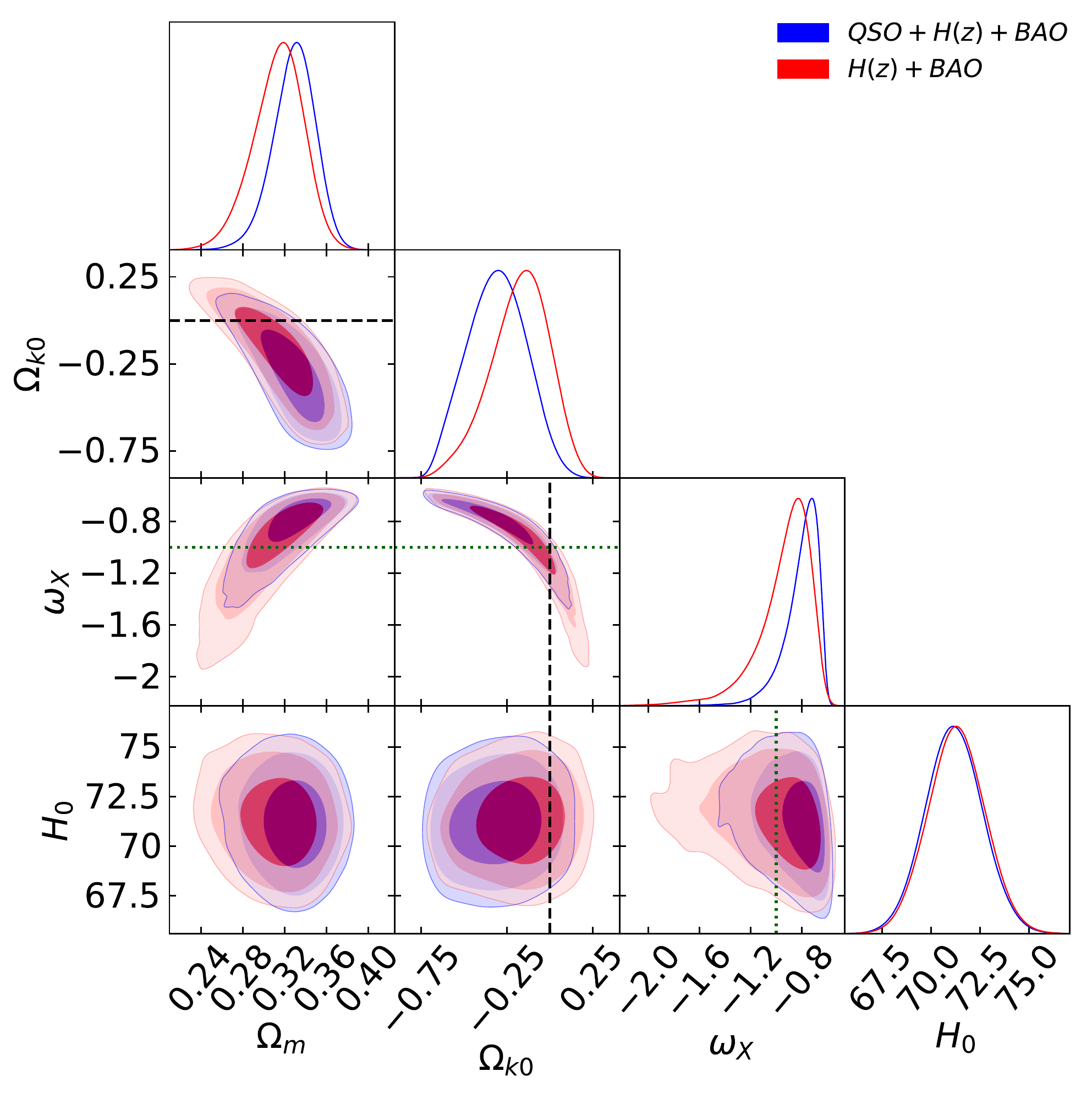}\par
\end{multicols}
\caption{Non-flat XCDM parametrization constraints from QSO (grey), $H(z)$ + BAO (red),  and QSO + $H(z)$ + BAO (blue) data. Left panel shows 1, 2, and 3$\sigma$ confidence contours and one-dimensional likelihoods for all free parameters. The red dotted curved lines in the $\omega_{K0} - \om$, $\omega_X - \om$, and $\omega_X - \Omega_{k0}$
panels are the zero acceleration lines with currently accelerated cosmological expansion occurring below the lines. Each of the three lines are computed with the third parameter set to the QSO data only best-fit value of Table 4. Right panel shows magnified plots for only cosmological parameters $\om$, $\ok$, $\omega_X$, and $H_0$, without the QSO-only constraints. These plots are for the $H_0 = 73.24 \pm 1.74$ ${\rm km}\hspace{1mm}{\rm s}^{-1}{\rm Mpc}^{-1}$ prior. The black dashed straight lines and the green dotted straight lines are $\ok$ = 0 and $\omega_x$ = $-1$ lines.}
\label{fig:non-flat XCDM68 model with BAO, H(z) and QSO data}
\end{figure*}
\begin{figure*}
\begin{multicols}{2}
    \includegraphics[width=\linewidth]{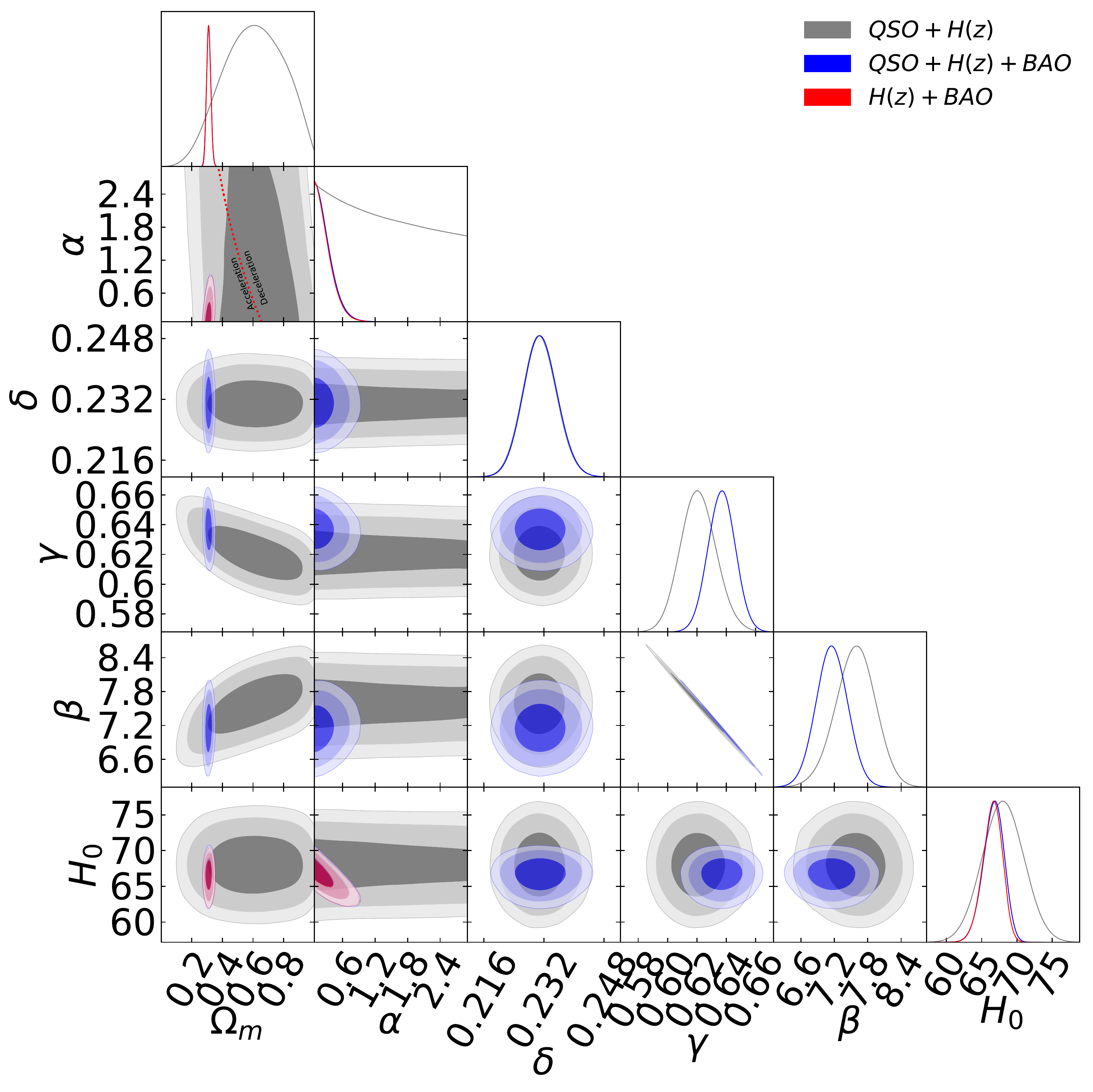}\par
    \includegraphics[width=\linewidth]{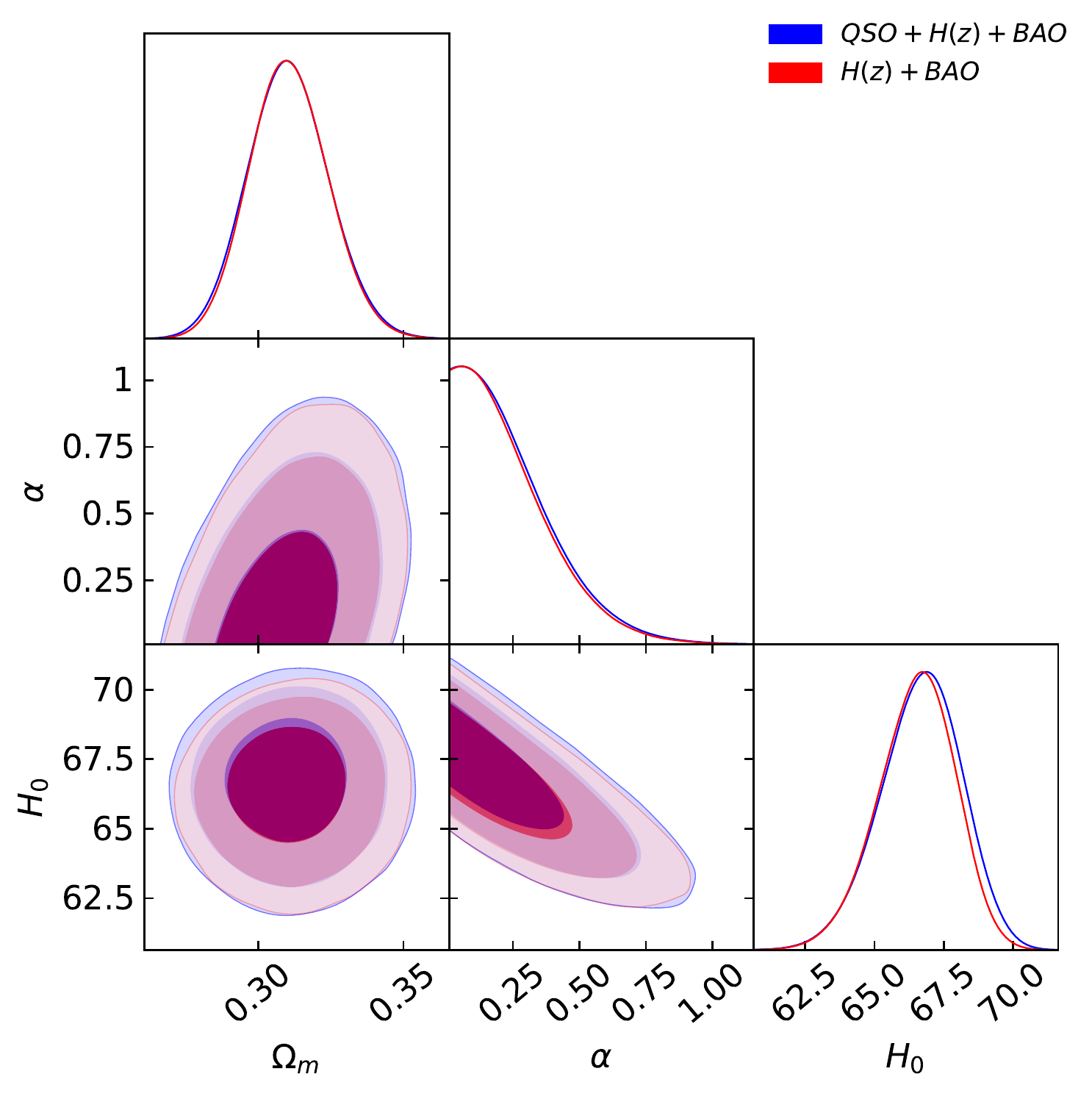}\par
\end{multicols}
\caption{Flat \pcdm\ model constraints from QSO (grey), $H(z)$ + BAO (red),  and QSO + $H(z)$ + BAO (blue) data. Left panel shows 1, 2, and 3$\sigma$ confidence contours and one-dimensional likelihoods for all free parameters. The red dotted curved line in the $\alpha - \om$ panel is the zero acceleration line, with currently accelerated cosmological expansion occurring to the left of the line. Right panel shows magnified plots for only cosmological parameters $\om$, $\alpha$, and $H_0$, without the QSO-only constraints. These plots are for the $H_0 = 68 \pm 2.8$ ${\rm km}\hspace{1mm}{\rm s}^{-1}{\rm Mpc}^{-1}$ prior.}
\label{fig:flat fphiCDM68 model with BAO, H(z) and QSO data}
\end{figure*}
\begin{figure*}
\begin{multicols}{2}
    \includegraphics[width=\linewidth]{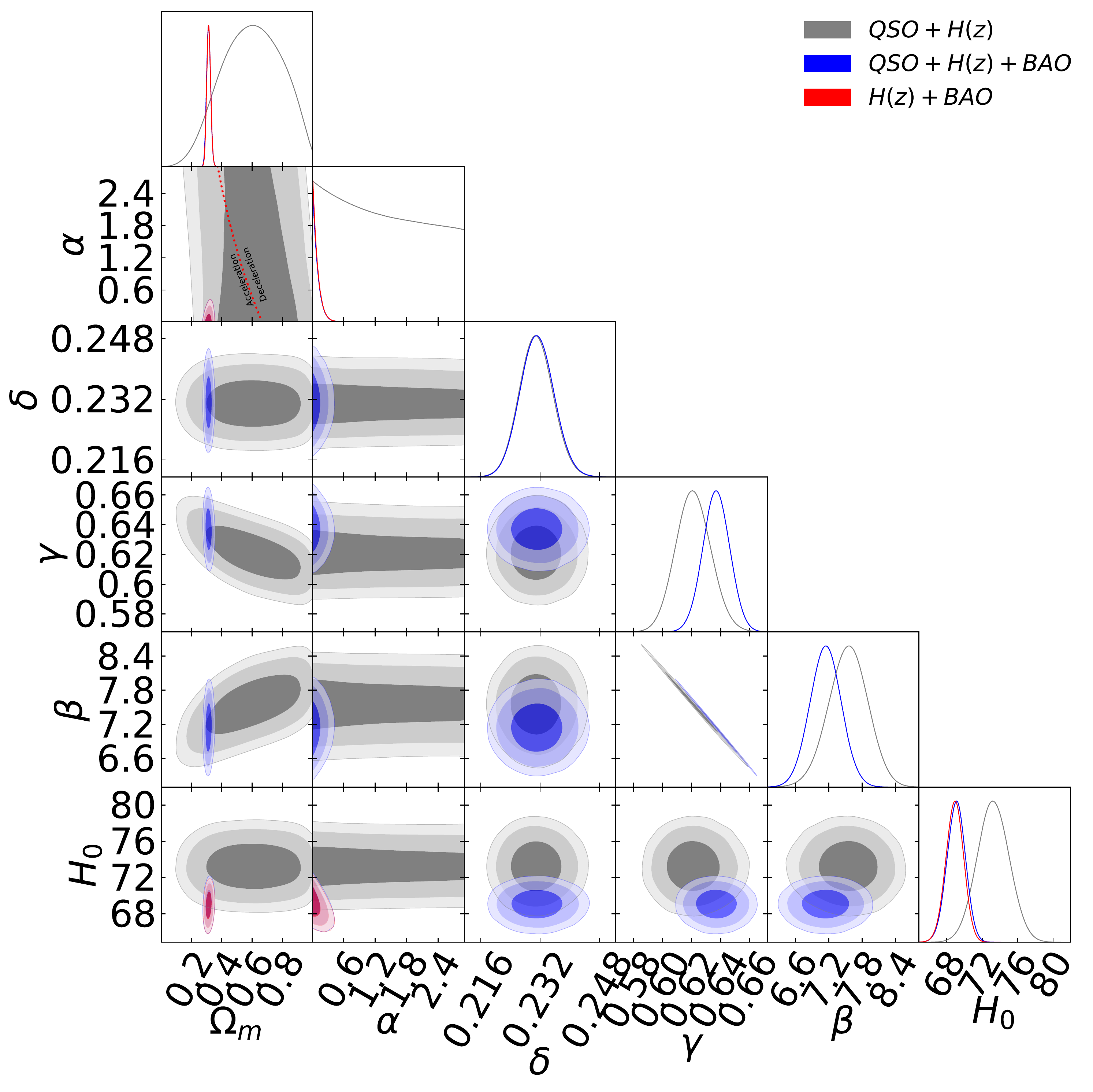}\par
    \includegraphics[width=\linewidth]{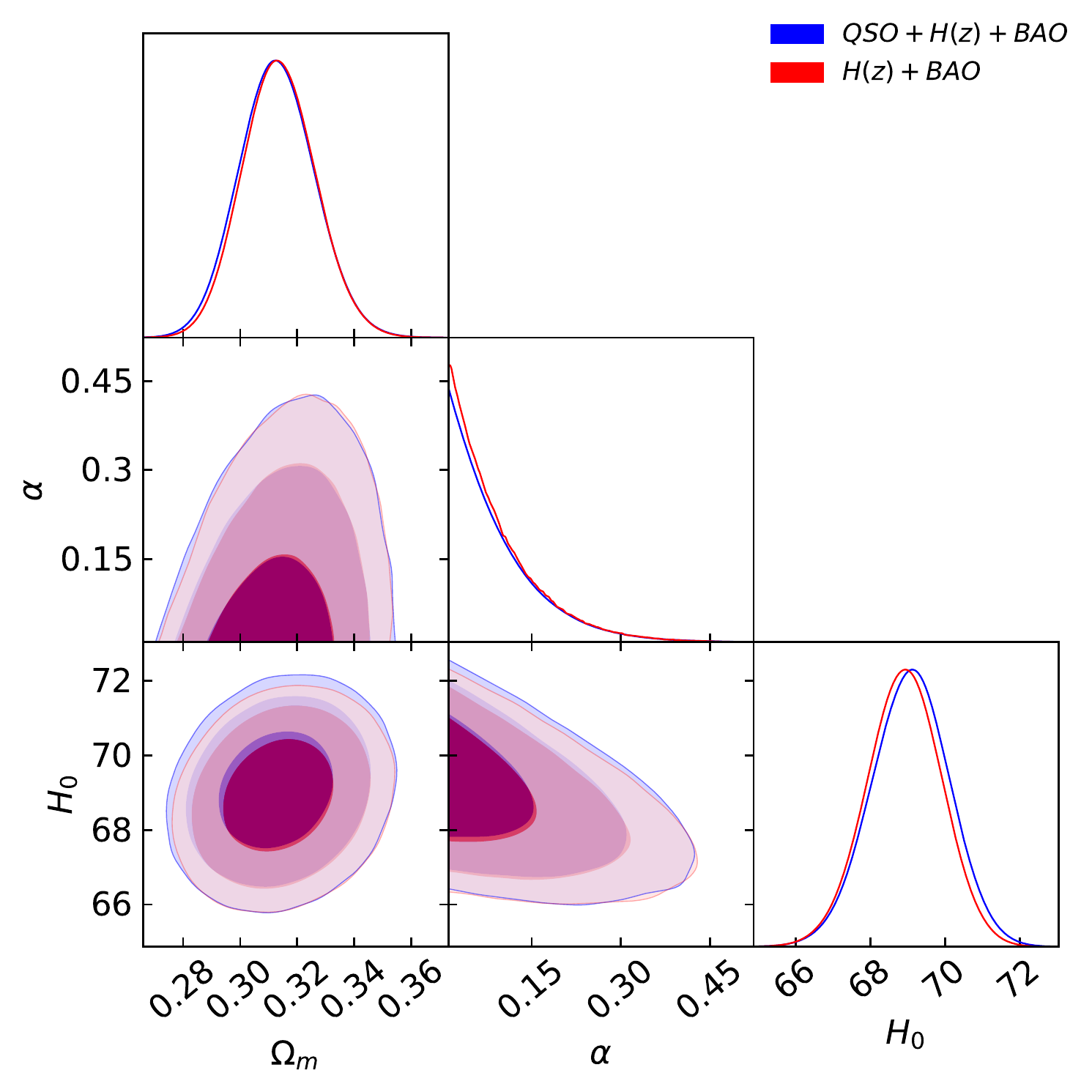}\par
\end{multicols}
\caption{Flat \pcdm\ model constraints from QSO (grey), $H (z)$ + BAO (red),  and QSO + $H(z)$ + BAO (blue) data. Left panel shows 1, 2, and 3$\sigma$ confidence contours and one-dimensional likelihoods for all free parameters. The red dotted curved line in the $\alpha - \om$ panel is the zero acceleration line, with currently accelerated cosmological expansion occurring to the left of the line. Right panel shows magnified plots for only cosmological parameters $\om$, $\alpha$, and $H_0$, without the QSO-only constraints. These plots are for the $H_0 = 73.24 \pm 1.74$ ${\rm km}\hspace{1mm}{\rm s}^{-1}{\rm Mpc}^{-1}$ prior.}
\label{fig:flat fphiCDM73 model with BAO, H(z) and QSO data}
\end{figure*}
\begin{figure*}
\begin{multicols}{2}
    \includegraphics[width=\linewidth]{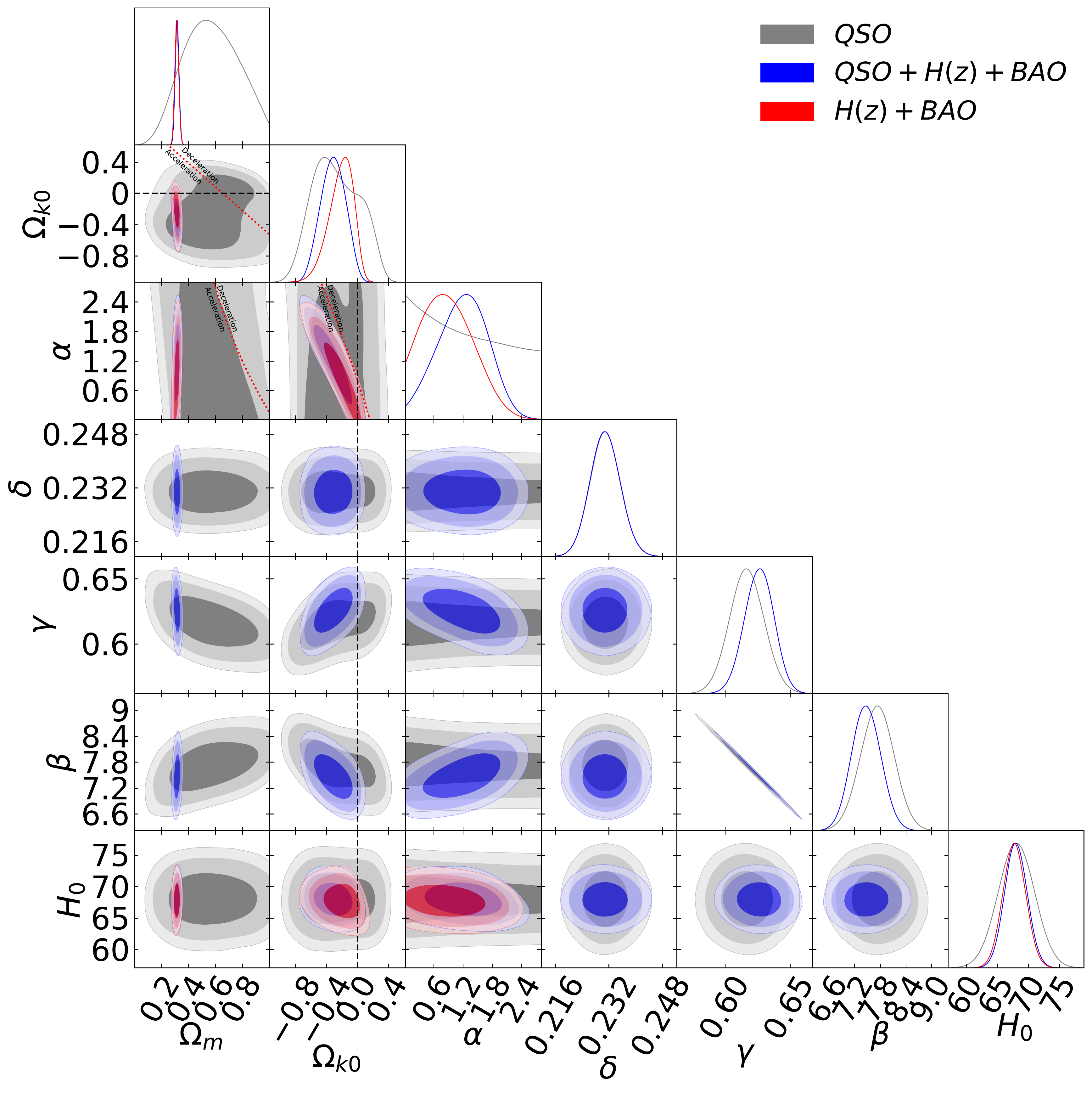}\par
    \includegraphics[width=\linewidth]{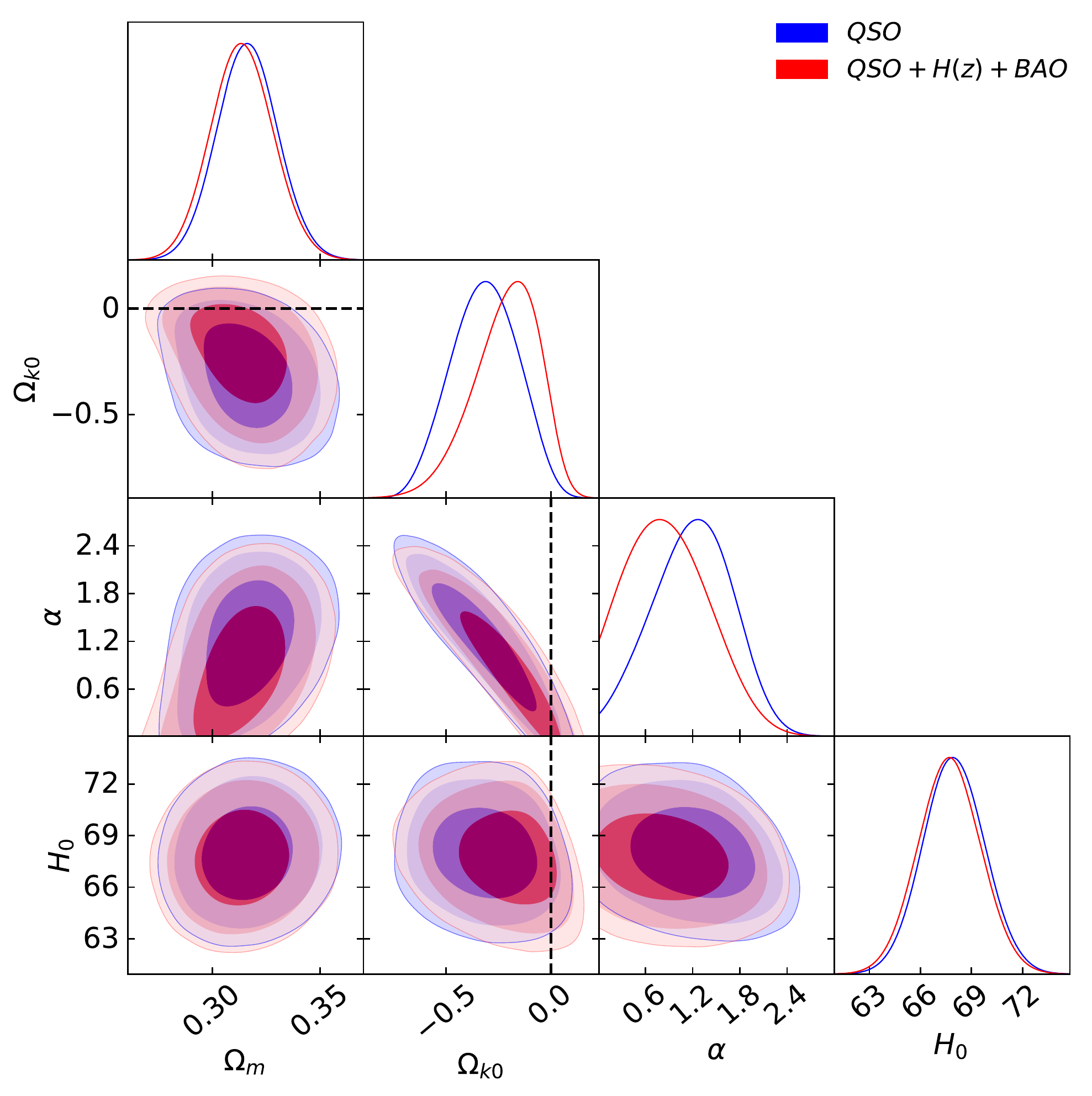}\par
\end{multicols}
\caption{Non-flat \pcdm\ model constraints from QSO (grey), $H(z)$ + BAO (red),  and QSO + $H(z)$ + BAO (blue) data. Left panel shows 1, 2, and 3$\sigma$ confidence contours and one-dimensional likelihoods for all free parameters. The red dotted curved lines in the $\omega_{K0} - \om$, $\alpha - \om$, and $\alpha - \Omega_{K0}$
panels are the zero acceleration lines with currently accelerated cosmological expansion occurring below the lines. Each of the three lines are computed with the third parameter set to the QSO data only best-fit value of Table 3. Right panel shows magnified plots for only cosmological parameters  $\om$, $\ok$, $\alpha$, and $H_0$, without the QSO-only constraints. These plots are for the $H_0 = 68 \pm 2.8$ ${\rm km}\hspace{1mm}{\rm s}^{-1}{\rm Mpc}^{-1}$ prior. The black dashed straight lines are $\ok$ = 0 lines.}
\label{fig: nfphiCDM73 model with BAO, H(z) and QSO data}
\end{figure*}
\begin{figure*}
\begin{multicols}{2}
    \includegraphics[width=\linewidth]{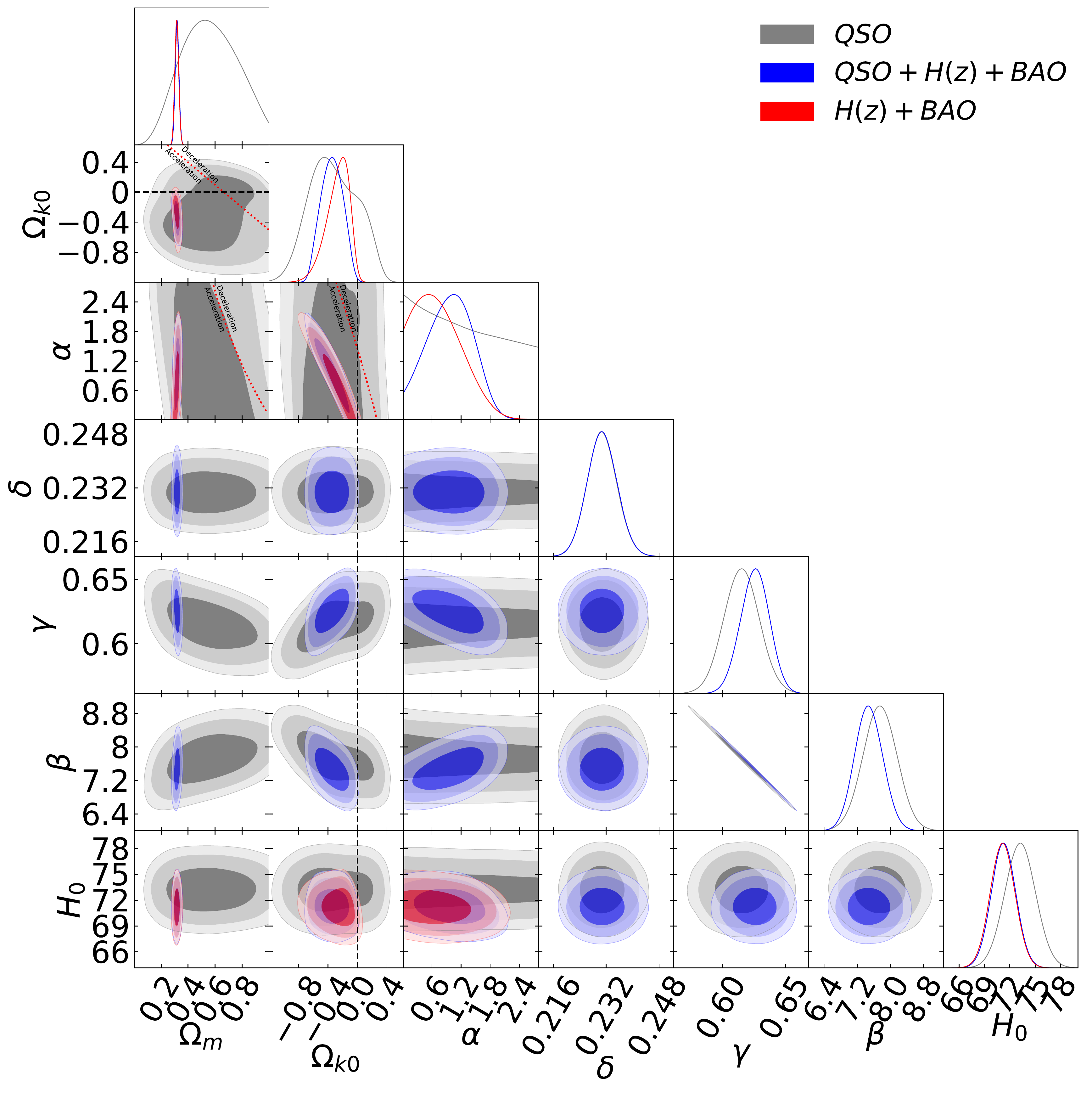}\par
    \includegraphics[width=\linewidth]{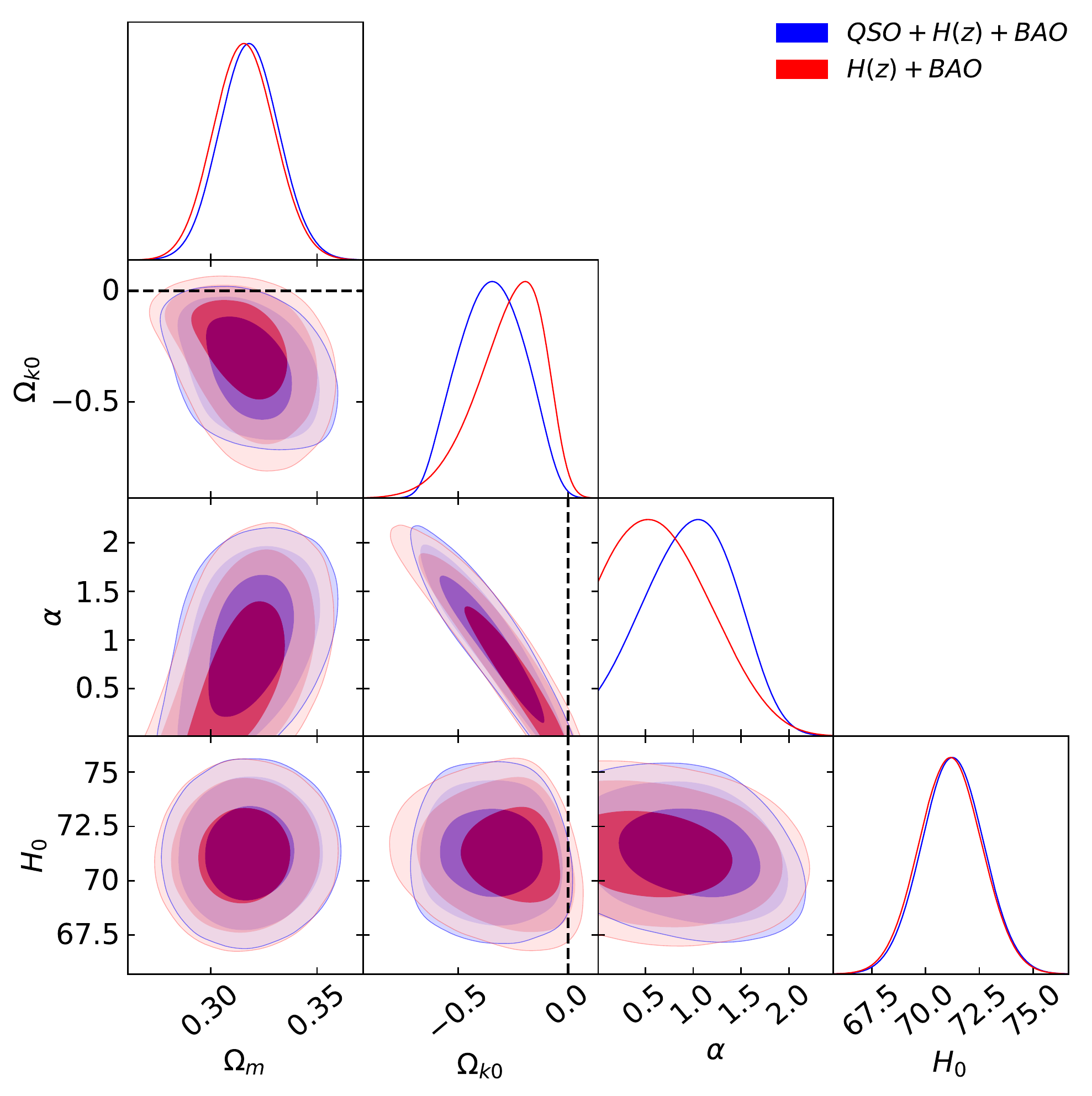}\par
\end{multicols}
\caption{Non-Flat \pcdm\ model constraints from QSO (grey), $H(z)$ + BAO (red),  and QSO + $H(z)$ + BAO (blue) data. Left panel shows 1, 2, and 3$\sigma$ confidence contours and one-dimensional likelihoods for all free parameters. The red dotted curved lines in the $\omega_{K0} - \om$, $\alpha - \om$, and $\alpha - \Omega_{K0}$
panels are the zero acceleration lines with currently accelerated cosmological expansion occurring below the lines. Each of the three lines are computed with the third parameter set to the QSO data only best-fit value of Table 4. Right panel shows magnified plots for only cosmological parameters $\om$, $\ok$, $\alpha$, and $H_0$, without the QSO-only constraints.These plots are for the $H_0 = 73.24 \pm 1.74$ ${\rm km}\hspace{1mm}{\rm s}^{-1}{\rm Mpc}^{-1}$ prior. The black dashed straight lines are $\ok$ = 0 lines.}
\label{fig: nfphiCDM73 model with BAO, H(z) and QSO data}
\end{figure*}

\bsp	
\label{lastpage}
\end{document}